\documentclass[aoas,MSNbibl,nameyear,seceqn,dvips]{arximspdf}
\usepackage{dcolumn}
\usepackage{graphicx}

\doi{10.1214/10-AOAS450}
\volume{5}
\issue{2B}
\pubyear{2011}
\firstpage{1488}
\lastpage{1511}

\makeatletter
\newcolumntype{d}[1]{D{.}{.}{#1}}
\newcommand{\aJ}{{\alpha_J}}
\newcommand{\mJ}{{\mu_J}}
\newcommand{\DV}{^\nabla}
\newcommand{\HP}{^{\mathrm{HP}}}
\newcommand{\PM}{^{\mathrm{PM}}}
\newcommand{\RR}{\varrho} % peak resolution
\newcommand{\bY}{\mathbf{Y}}
\newcommand{\Yb}{{\bar Y}}
\newcommand{\bb}{\beta_{0}(\mz)} % bck ground
\newcommand{\bc}{S} % bck ground constant
\newcommand{\con}{\zeta} % scale conston
\newcommand{\dT}{T} % Could be [\ppb-\ppa] or {\Delta T}
\newcommand{\gamR}{\lambda} % gamma process rate parameter
\newcommand{\hh}{\eta} % peak ht
\newcommand{\kk}{{k}} % kernel
\newcommand{\mnY}{\mu }
\newcommand{\mz}{{t}}
\newcommand{\nLsr}{l}
\newcommand{\nPks}{J}
\newcommand{\obs}{^{\mathsf{ob}}}
\newcommand{\pp}{\tau} % peak loc
\newcommand{\ppa}{{T_0}} % TIME min
\newcommand{\ppb}{{T_1}} % TIME max
\newcommand{\pr}{\varphi} % precision
\newcommand{\sgY}{f}
\newcommand{\sqrtlf}{\sqrt{\log4}} % Sqrt[log 4]
\newcommand{\thb}{{\bolds\theta}}
\newcommand{\ww}{\omega} % peak width
\newcommand{\Be}{{\mathsf{Be}}}
\newcommand{\Ga}{{\mathsf{Ga}}}
\newcommand{\N}{{\mathsf{N}}}
\newcommand{\NB}{{\mathsf{NB}}}
\newcommand{\Un}{{\mathsf{Un}}}
\newcommand{\Eo}{\mathrm{E}_1{}}
\newcommand{\Eqn}[1]{equation~(\ref{#1})}
\newcommand{\E}{\mathsf{E}}
\newcommand{\Fig}[1]{Figure~\ref{#1}}
\renewcommand{\P}{\mathsf{P}}
\newcommand{\Sec}[1]{Section~\ref{#1}}
\newcommand{\Tbl}[1]{Table~\ref{#1}}
\newcommand{\bbR}{\mathbb{R}}
\newcommand{\eps}{\epsilon}
\renewcommand{\epsilon}{\varepsilon}
\newcommand{\etc}{etc}
\newcommand{\half}{\tfrac12}
\newcommand{\iN}{_{1\le i\le n}}
\newcommand{\iid}{\stackrel{\mathrm{i.i.d.}}{\sim}}
\newcommand{\ind}{\stackrel{\mathrm{ind}}{\sim}}
\newcommand{\jJ}{_{1\le j\le J}}
\newcommand{\one}[1]{\mathbf1_{\{#1\}} }
\renewcommand{\mid}{|}

\def\bsuffix #1{#1}

\def\@bmisc[#1]{%
  \get@battribute{unstr}%
  \common@pub@types%
  \let\bauthor\bbl@bauthor%
  \let\bhowpublished\@firstofone%
  \def\borganization##1{{\bauthor@style ##1}}%
}

\makeatother

\begin{document}
\begin{frontmatter}

\title{Bayesian nonparametric models for peak identification
in MALDI-TOF mass spectroscopy\thanksref{TT1}}
\vspace*{3pt}
\runtitle{LARK models for peak identification}

\begin{aug}
\author[A]{\fnms{Leanna~L.}~\snm{House}\thanksref{t2}\ead[label=e1]{lhouse@vt.edu}},
\author[B]{\fnms{Merlise~A.}~\snm{Clyde}\corref{}\thanksref{t2,t3}\ead[label=e2]{clyde@stat.duke.edu}}
\and
\author[B]{\fnms{Robert~L.}~\snm{Wolpert}\thanksref{t4}\ead[label=e3]{rlw@stat.duke.edu}}
\vspace*{3pt}
\runauthor{L. L. House, M. A. Clyde and R. L. Wolpert}
\affiliation{Virginia Tech, Duke University and Duke University}
\address[A]{L. L. House\\
Department of Statistics \\
Virginia Tech\\
Blacksburg, Virginia 24061-0439\\
USA\\
\printead{e1}} %adresu isvedimo komanda gale!
\address[B]{M. A. Clyde\\
R. L. Wolpert \\
Department of Statistical Science\\
Duke University\\
Durham, North Carolina 27708-0251\\ USA\\
\printead{e2}\\
\phantom{\textsc{E-mail:}\ }\printead*{e3}}
\end{aug}
\thankstext{TT1}{Supported by  NSF   Grant   DMS-04-22400.
Any opinions, findings, and conclusions or recommendations expressed
in this material are those of the authors and do not necessarily
reflect the views of the National Science Foundation.}
\thankstext{t2}{Supported in part by NSF
Grant DMS-03-42172.}
\thankstext{t3}{Supported in part by NSF
Grant DMS-04-06115 and NIH Grant R01-HL090559-01.}
\thankstext{t4}{Supported in part by NSF Grants DMS-07-57549 and
PHY-09-41373, and NASA Grant NNX09AK60G.}

% HISTORY:
\received{\smonth{7} \syear{2010}}
\revised{\smonth{11} \syear{2010}}

% ABSTRACT
%
\vspace*{5pt}
\begin{abstract}
We present a novel nonparametric Bayesian approach based on L{\'{e}}vy
Adaptive Regression Kernels (LARK) to model spectral data arising
from MALDI-TOF (Matrix Assisted Laser Desorption Ionization
Time-of-Flight) mass spectrometry. This model-based approach
provides identification and quantification of proteins through model
parameters that are directly interpretable as the number of
proteins, mass and abundance of proteins and peak resolution, while
having the ability to adapt to unknown smoothness as in wavelet based
methods. Informative prior distributions on resolution are key to
distinguishing true peaks from background noise and resolving broad
peaks into individual peaks for multiple protein species. Posterior
distributions are obtained using a reversible jump Markov chain
Monte Carlo algorithm and provide inference about the number of
peaks (proteins), their masses and abundance. We show through
simulation studies that the procedure has desirable true-positive
and false-discovery rates. Finally, we illustrate the method on
five example spectra: a~blank spectrum, a spectrum with only the
matrix of a low-molecular-weight substance used to embed target
proteins, a spectrum with known proteins, and a single spectrum and
average of ten spectra from an individual lung cancer patient.
\vspace*{5pt}
\end{abstract}

% KEYWORDS
%

\end{frontmatter}

%s1 ###
\section{Introduction} \label{s:intro}

Recent innovations in protein separation methods, ionization procedures and
detection algorithms have led mass spectrometry (MS) to play a vital
role in
the explosive growth of proteomics [\citet{Dass2001}, page xxi]. Despite
technological advances in data collection, it remains challenging to extract
biologically relevant information (such as biomarkers) from MS spectral data
[\citet{Dass2001}, Chapters 3 and 5;
\citet{CoomKoometal2005};
\citeauthor{BaggMorrCoom2004} (\citeyear{BaggMorrCoom2004,BaggCoomMorr2006});
Clyde, House and Wol\-pert \citeyear{ClydHousWolp2006};
\citet{MorrBrowetal2006}].

Identifying peak locations (which represent proteins) and quantifying
protein abundance in spectra is often preceded by a multi-stage
analysis involving calibration, normalization, baseline subtraction
and filtering of noise [\citet{MorrCoometal2005};
\citet{Tibsetal2004};
\citet{YasuMcLeetal2003}]. A problem with such an approach is that each
individual step may introduce errors, artifacts or biases that may
interfere with later stages of the analyses such as classification of
subjects or identification of biomarkers. Methods that model
background, noise and features simultaneously may lead to improved
classification or inferences [\citet{CoomTsavetal2005}].
Nonparametric methods such as wavelet regression have proved
successful in simultaneously modeling background and denoising,
allowing one to extract features or regions of spectra that
differentiate groups [\citet{YasuMcLeetal2003};
\citet{CoomTsavetal2005};
\citet{WangRayMall2007};
\citet{MorrBrowHerretal2008}]. While wavelets are
well suited for modeling local features like spectral peaks, neither
the scales and locations that index basis functions nor coefficients
used in the wavelet representation of expected intensity have any
inherent biological interpretation for the typical wavelets used in
practice, such as Daubechies' ``least asymmetric'' \texttt{la8}
wavelet family. As an alternative to nonparametric regression for
modeling intensities, \citet{GuinDoetal2006} and \citet
{MullBaggetal2010} developed a Bayesian mixture model based on beta
distributions to estimate a~density function for time-of-flight. The
parameters of this model are more interpretable than the wavelet
regression methods, but the approach does not incorporate information
about peak resolution, which we will show allows one to resolve broad
peaks into multiple peaks.

In this paper we propose a novel nonparametric method employing L\'{e}vy
Adaptive Kernel Regression (LARK) models [\citet{WolpClydTu2006};
\citet{ClydWolp2007}], which retains the adaptivity and flexibility that make
wavelet and other nonparametric methods appealing but, in contrast to these
other methods, uses model parameters with direct biological interpretations.
This offers the opportunity to elicit meaningful expert opinion to
guide the
selection of prior distributions, and  a posteriori  to provide
posterior distributions on model parameters that are meaningful to the
expert. The model presented in this article is intended for use with a
``single'' spectrum, although an entire set of exchangeable spectra may be
analyzed by applying the method to the mean spectrum, similar to the
mean-spectrum undecimated wavelet thresholding (MUDWT) method of
\citet{MorrCoometal2005}.

The paper is arranged as follows. We begin in \Sec{s:MS} with a brief
overview of MALDI-TOF mass spectrometry. In \Sec{s:MOD} we develop a
statistical model for protein abundance as a function of
time-of-flight based on the recently developed nonparametric LARK
models. Prior distributions for the model parameters are developed in
\Sec{s:PRI} from elicited expert knowledge about the MALDI-TOF
procedure and from exploratory analysis of MALDI-TOF data from related
experiments. Inference about parameters of clinical interest are
obtained from posterior distributions and are described in \Sec
{s:POS}. We then illustrate the methodology in \Sec{s:imp} using a
series of experiments with real data. We use laboratory data from
three experiments with known sources: blank spectra, which help to
characterize noise in MALDI-TOF experiments; matrix spectra, which
help to characterize background; and a known protein mixture, which
illustrates the model's ability to resolve masses. The LARK model is
also applied to single and mean spectra from a recent lung cancer
study conducted at the Duke University Medical Center. In \Sec{s:sim}
we the compare the LARK model to the wavelet method of
\citet{MorrCoometal2005} using mean spectra from the TOF simulator
of \citet{CoomTsavetal2005}. We conclude with a discussion and
suggestions for future work in \Sec{s:dis}.

%s2 ###
\section{MALDI-TOF data}\label{s:MS}

In Matrix Assisted Laser Desorption Time-of-Flight Mass Spectrometry, or
MALDI-TOF MS, inference about the molecular composition of a sample is based
on indirect measurement of molecular masses. Molecules are initially
embedded in a \textit{matrix} of low molecular weight substance, such as
sinapinic acid, and placed on a metal plate. The molecules are then
simultaneously dislodged (by vaporizing the substrate) and ionized (by
removing one or more electrons from molecules) by a series of laser pulses.
The now-charged molecules are accelerated by a strong electric field
toward a
detector. In the Applied Biosystems Voyager DE Biospectrometry Workstation
[\citet{UsrGde-Voy}],
% is a linear MALDI-TOF Delayed Extraction mass spectrometer. Voyager's
%
a linear detector measures ion abundance over time, then sends a signal at
regular time intervals (\textit{clock ticks}) to a digitizer for
conversion to
measured intensities; for the examples considered below, samples are
taken at
regular $4$~ns or $16$~ns intervals. The reported intensity in each time
interval is typically the aggregate sum over several repeated laser
``shots,'' leading to what we will refer to as a~single spectrum, with
response ion intensity at corresponding time-of-flights (TOFs).

\begin{figure}

\includegraphics{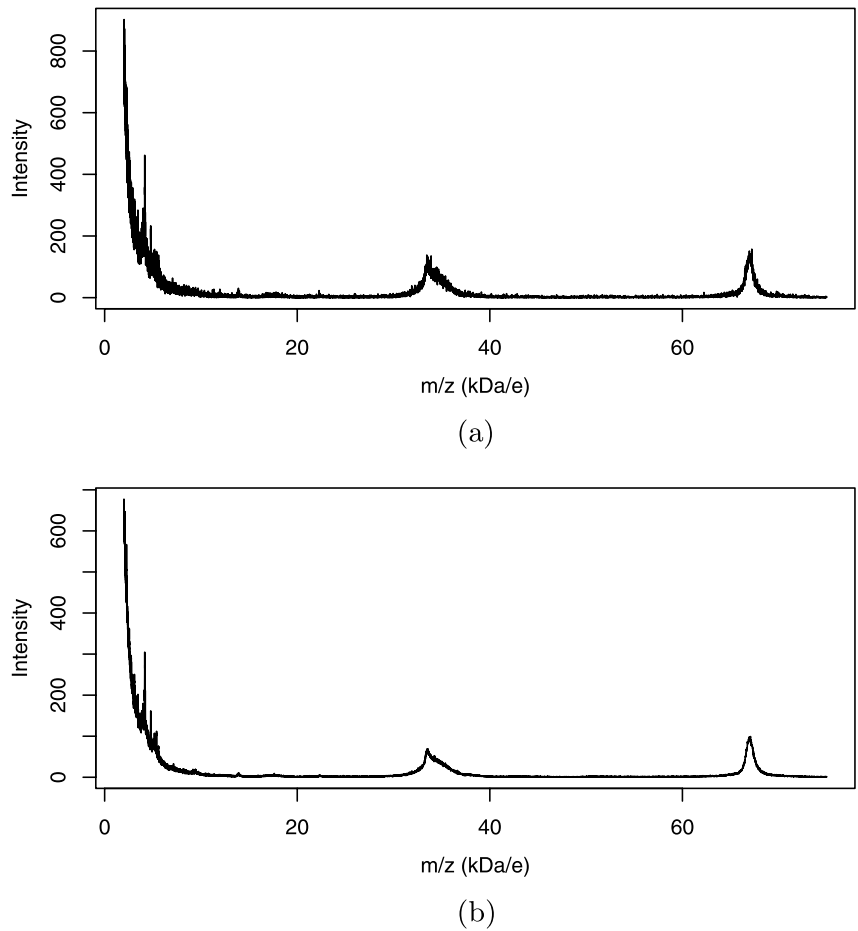}

\caption{Single spectrum \textup{(a)} and mean of ten spectra \textup{(b)} from a lung cancer
patient.}\label{fig:MS}\vspace*{-1.5pt}
\end{figure}

\Fig{fig:MS} illustrates serum protein spectra from a single
individual with
lung cancer from a study conducted at the Duke Medical Center Radiology
Department [\citet{WangHowaCampEtal2003}].
%%CampWangHowaEtal2003, "1652--1656",
%%Camp:Fitz:Patz:2003, "1659--1660",
%%Howa:Wang:Camp:Etal:2003, "1720--1724",
Each serum sample was separated into $20$ fractions along a pH
gradient prior to the MALDI-TOF analysis to reduce saturation of the
signal. Ten replicated spectra were obtained for each fraction, each
with ten laser shots, using Voyager with a sinapinic acid matrix. For
our analyses we randomly selected one fraction from one subject;
\Fig{fig:MS} shows the single spectrum from the chosen
subject-fraction~(a), and the mean spectrum obtained by averaging the
intensities of all ten replicates (b) from the same fraction.
\citet{MorrCoometal2005} suggest using the mean spectrum as a way
to reduce noise from various sources as discussed below.
%

%
%f1 ###

Distance traveled under constant acceleration is a quadratic function of
time, leading to a simple but nonlinear relationship between TOF and the
molecules' masses and ionic charge (the latter two enter only through their
quotient, the \textit{mass to charge ratio} m$/$z [\citet{CoomKoometal2005}]).
Under ideal conditions the TOF spectrum would show a spike at the TOF
corresponding to each molecular species present.
In actual MALDI-TOF spectra (\Fig{fig:MS}) we observe irregular peaks rather
than one-dimensional spikes because molecules of equal size and charge
do not
all reach the detector at the same time. The most important of the many
causes of TOF dispersion is variability in the amount of ionizing laser
energy received by molecules of varying location within the matrix; those
further from the matrix surface or from the center of the laser pulse may
receive less kinetic energy and thus have lower initial velocities than
similarly-sized molecules located closer to the center, delaying their
arrival at the detector. Molecules may exchange energy in collisions, and
may lose or gain mass through fragmentation and agglomeration, respectively.
All these lead to TOF variation for each molecular species [\citet
{CoomKoometal2005};
\citet{ZhiGarr1998};
\citet{Fran1997}]. Furthermore, while the abundance
of a protein in the sample ideally corresponds to the area under the TOF
distribution curve, factors such as ion suppression, multiple charged ions,
adducts (the addition of other molecules to the protein), protein--protein
interactions and isotope distributions may result in a protein being
represented by more than one peak in the spectrum.\vadjust{\goodbreak}

The interpretation and analysis of MALDI-TOF data are complicated by several
other sources of variation described by \citet{MorrCoometal2005} and
\citet{CoomKoometal2005}. In addition to \textit{measurement error} which
may mask or distort protein peaks, at least three other sources complicate
the comparison or synthesis of multiple spectra: \textit{calibration}
(uncertainty in the conversion of TOF to m$/$z, including variable latency
that affects time registration); \textit{background} (a constant or even
time-varying trend in the overall level); and \textit{scale} (caused by many
things including variability of laser intensity). One way to accommodate
this is to construct models for peak identification and quantification that
incorporate these recognized sources of variability, as in the wavelet
approach of \citet{CoomTsavetal2005},
\citet{MorrCoometal2005} developed for
calibrated spectra. Our approach, based on kernel models, has the added
advantage that model parameters have direct physical interpretations.

%s3 ###
\section{A model for MALDI-TOF}\label{s:MOD}

To reduce the variability attributable to differing numbers of laser shots
and differing baselines, we first standardize the spectrum and model
% the standardized spectrum at TOF $\mz$, for some range $\ppa\le\mz
% \ppb$,
%
%e3.1 ###
\begin{equation}\label{eqn:tr}
Y_\mz\equiv\frac{Y\obs_\mz- \min(\bY\obs)}{\nLsr}
%= \big[Y\obs_\mz- \min(\bY\obs)\big]/\nLsr
\end{equation}
for a raw spectrum $\bY\obs\equiv\{Y\obs_t\}$ for ${\ppa\leq t
\leq
\ppb}$, where $\ppa$ and $\ppb$ correspond to the range of TOFs of
scientific interest and $l$ is the number of laser shots summarized by
$\bY\obs$. The initial molecular velocities are expected to be approximately
Gaussian in distribution [\citet{Dass2001}, page 75]. This and the
dynamics of
the MALDI-TOF process [\citet{CoomKoometal2005}] suggest that TOFs for
a single
isotopic peak will also have symmetric bell-shaped distributions in the time
domain, leading us [and others---see \citet{MorrCoometal2005};
\citet{Malyetal2005}] to prefer time of flight (TOF, in $\mu${s}) rather than
mass-to-charge ratios (m$/$z, in Da$/$e) for spectral modeling
(although we
follow convention in reporting and plotting results results in m$/$z).
Because ions from the matrix may saturate the detector at initial TOFs
[\citet{Malyetal2005}] and masses less than $2$~kDa were not of scientific
interest to our collaborators, $\ppa$ will correspond to the TOF of a
mass of
2~kDa throughout, unless otherwise noted. While the nonparametric model
that we propose can accommodate an arbitrary lower bound (even $\ppa= 0$),
modeling these extra initial peaks will increase the running time of the
algorithm, with little or no improvement in peak identification or
model fit
for the rest of the spectrum.

%s3.1 ###
\subsection{Peak shape}\label{ss:shape}

The shape of a symmetric isotopic peak may be represented by a probability
density function for TOF $\mz$ with parameters governing the protein peak's
location $\pp$ and width $\ww$. Examples include the Gaussian
%
%e3.2 ###
\begin{equation}
\label{eq:gauss}
\kk(\mz;\pp, \ww) = \frac1{\sqrt{2\pi} \ww} \exp( {-|\mz-\pp
|^2/2\ww^2} )
\end{equation}
and Cauchy (sometimes called Lorentzian in the MS literature)
%
%e3.3 ###
\begin{equation}\label{eqn:cau}
\kk(\mz; \pp, \ww) = \frac{\ww}{\pi(\ww^2+|\mz-\pp|^2)},
\end{equation}
as in \citet{Dass2001}, page 75, \citet{Kempetal2004}, and
\citet{UsrGde-Voy}, pages 6--30.

A protein signature associated with $\nPks$ peaks may now be
represented as a sum
%
%e3.4 ###
\begin{equation}\label{eqn:sig}
\sgY(\mz)=\sum_{j=1}^{\nPks} \kk(\mz;\pp_j,\ww_j)  \hh_j,
\end{equation}
where $\{\pp_j\}$, $\{\ww_j\}$ and $\{\hh_j\}$ represent the unknown
location (TOF), peak width and abundance of the $j$th protein (or
molecule), respectively. This is a special case of the L\'{e}vy Adaptive
Regression Kernel (LARK) models of
\citet{WolpClydTu2006},
\citet{ClydWolp2007}, which generalize classical
kernel regression [\citet{WandJone1995}] by allowing the number of
kernels and the ``smoothing parameters'' ($\ww)$ of the kernel
$k$ to adapt to the unknown degree of smoothness in the data, as
in wavelet models [\citet{MorrCoometal2005}]. While both methods
lead to excellent function reconstructions, the parameters in the
kernels $(\pp_j, \ww_j)$ and kernel coefficients $\hh_j$ of the LARK
model have direct biological interpretations which aide in prior
specification (detailed below) and posterior interpretation.

%s3.2 ###
\subsection{Peak width and resolution}\label{ss:width}

\begin{figure}[b]

\includegraphics{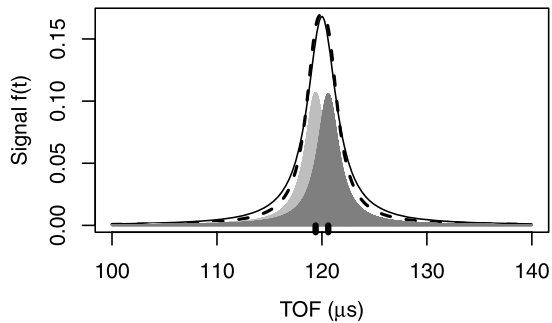}

\caption{The (nearly indistinguishable) lines represent
simulated protein signals from a~sample with either one wide peak (solid)
centered at $120$ $\mu${s} or a mixture (dashed) of two narrow peaks
centered at
bold ticks.}\label{fig:resolve}
\end{figure}

Protein peaks tend to be broader for late-arriving molecules than for earlier
ones, with width nearly proportional to arrival time [\citet
{Siuz2003}, page 44]; for this reason it is conventional in mass
spectrometry to
quantify the precision (narrowness) of a kernel $\kk( \cdot ;\pp
,\ww)$ not
by the scale $\ww$, but by the \textit{resolution}
%
%e3.5 ###
\begin{equation}
\label{eq:resolution-def}
\rho\equiv\pp/\Delta\pp,
\end{equation}
where $\Delta\pp$, the so-called \textit{full width at half mass} or
FWHM, is the width of the kernel $\kk( \cdot ;\pp,\ww)$ at half its
height [\citet{Dass2001}, page 120]. For a symmetric kernel, $\Delta
\pp$ is the solution of the equation
\[
\kk\bigl(\pp\pm\half\Delta\pp; \pp,\ww\bigr) = \half\kk(\pp; \pp,\ww
).
\]
For the
Gaussian and Cauchy kernels we have $\Delta\pp=2\ww\sqrtlf$ and
$\Delta\pp=2\ww$, respectively, leading to $\ww=\ww(\pp,\rho)$ with
%
%e3.6 ###
\begin{equation}\label{eqn:rr2ww}
\ww(\pp,\rho) = \frac{\pp}{2\rho{\sqrtlf}} \quad\mbox
{and}\quad
\ww(\pp,\rho) = \frac{\pp}{2\rho},
\end{equation}
for the Gaussian and Cauchy kernels, respectively. Prior knowledge about resolution can be used to resolve the ambiguity
illustrated in \Fig{fig:resolve}, where the observed spectrum may
arise from
either a single wide peak or a pair of nearby narrower peaks. As depicted
later in \Fig{fig:lung}(d)--(f), we illustrate how the model is able to
``deconvolve'' a wide peak into several individual protein peaks in real
data.

%f2 ###

%s3.3 ###
\subsection{Background noise sources}\label{ss:bkgd}

Even in the absence of any protein mole\-cules [i.e., with
$\sgY(\mz)\equiv0$], the MALDI-TOF spectrum does not vanish.
\Fig{fig:blank}(a) shows the nearly-constant level of thermal noise
from a
run with an empty plate, while \Fig{fig:matrix}(b) illustrates the
rapidly-decreasing signal with only the sinapinic acid matrix, showing the
early arrival at the detector of ionized matrix molecules (far lighter than
typical proteins under study). Since the signal from matrix ions dominates
the thermal noise or detector ``ringing,'' together these sources contribute
a background that falls off nearly exponentially to a nonzero asymptote.

Exploratory analysis suggests that the matrix molecular signal $\bb$ can
be modeled adequately as a constant (see below) plus an exponential function,
%
%e3.7 ###
\begin{equation}\label{eqn:EK}
\bb= \kk_0(\mz;\ww_0) \hh_0
= \frac{\hh_0}{\ww_0}\exp\{{-\mz/\ww_0}\}\one{\mz>0},
\end{equation}
with a characteristic decay time of $\ww_0>0$ and intensity $\hh_0>0$.

%s3.4 ###
\subsection{Expected spectral intensity}\label{ss:mean}

To reflect all of these features, we model the expected spectral
intensity as
%
%e3.8 ###
\begin{equation}
\mnY(\mz) = \con \{(1-\bc) +\bc [\sgY(\mz) + \bb
] \}
\label{eqn:sum}
\end{equation}
for an overall scale $\con$, a dimensionless signal-to-background ratio
$\bc\in[0,1]$, the protein signal $\sgY(\mz)$ from \Eqn{eqn:sig},
and the
matrix molecular signature $\bb$ from \Eqn{eqn:EK}. The term $\bc$
represents the proportion of observed intensity produced by molecular signal
(both matrix and protein), rather than by thermal noise.

%s3.5 ###
\subsection{Likelihood}\label{ss:lh}

\begin{figure}

\includegraphics{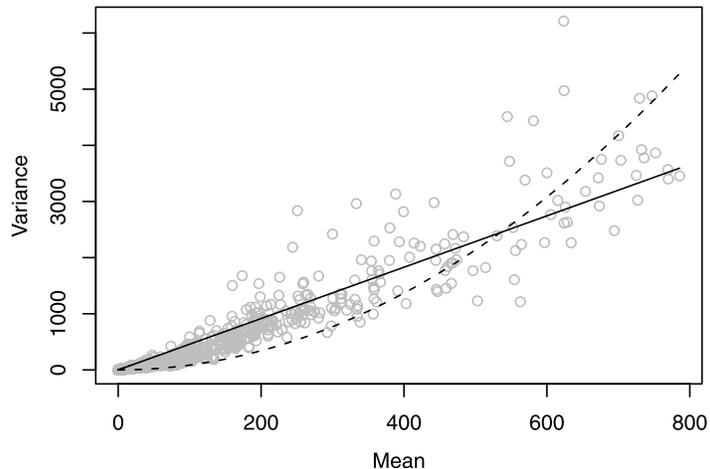}

\caption{Linear
(solid) and quadratic (dashed) fits of variance versus mean intensity of
intensity for $200$ $\mu${s} blocks of observations from a single spectrum
from a
lung cancer patient.}\label{fig:rlshp}
\end{figure}

Both gamma and log-normal distributions are commonly used to model positive
data like the standardized responses $Y_\mz$. The variance is proportional
to the mean for gamma distributions and to the square of the mean for
log-normals.
Exploratory data analysis (from both a Box--Cox approach and a robust
regression illustrated in \Fig{fig:rlshp}) suggests that the conditional
variance of standardized MS data $Y_\mz$, given the mean, is nearly
proportional to the first power of the mean, supporting the gamma model
%
%e3.9 ###
\begin{equation}\label{eqn:model}
Y_{\mz}\mid{\mnY( \cdot ), \pr}\ind\Ga(\pr\mnY(\mz), \pr),
\end{equation}
%
%mean:variance ratio
with mean $\mnY(\mz)$ and relative precision parameter $\pr$ (similar
relationships hold for the other data sets, although the slopes vary). This
leads to a~measure\-ment-error model with likelihood function
%
%e3.10 ###
\begin{equation}\label{eqn:LH}
\mathcal{L}(\thb;\bY) = \prod_{i=1}^n \Ga(Y_{\mz_i};\pr\mnY(\mz
_i), \pr)
\end{equation}
for the parameter vector $\thb$ comprising the conditional mean function
$\mnY( \cdot )$ [or, equivalently from \Eqn{eqn:sum}, all of $\con$,
$\nPks$, $\{\pp_j,\ww_j,\hh_j\}\jJ$, $\bc$, $\ww_0$, and~$\hh_0$] and~$\pr$.
Here $\bY=\{Y(\mz_i)\}\iN$ represents the %$n$-dimensional
vector of standardized intensities from (\ref{eqn:tr}), and
$\Ga(y;\alpha,\beta) =\frac{\beta^\alpha} {\Gamma(\alpha
)}y^{\alpha-1}
e^{-\beta y}\one{y>0}$ is the probability density function at $y\in
\bbR$ for
the gamma $\Ga(\alpha,\beta)$ distribution.

%
%f3 ###

Typically the likelihood function of \Eqn{eqn:LH} has many modes
because it
is difficult to distinguish wide peaks from clusters of narrow ones, or small
peaks from noise, from the data alone. Estimating $\thb$ (and in particular
$\nPks$, the number of protein peaks) by direct maximization of the
likelihood leads to over-fitting the data and to over-estimating $\nPks$.
This can be overcome by regularization or by a Bayesian approach, in which
prior distributions effectively penalize overly complex models.

%s4 ###
\section{Prior distributions for MALDI-TOF}\label{s:PRI}

We now address the problem of constructing a joint prior distribution
for all
the unknown parameters of the model of
\Sec{s:MOD},
%%
%Y_{\mz}\mid{\mnY( \cdot ), \pr}&\ind\Ga(\pr\mnY(\mz), \pr)\label{eqn:mod}\\
%%
%
%e4.1 ###
\begin{eqnarray}\label{eqn:mod}
Y_{\mz}\mid{\mnY( \cdot ), \pr}&\ind&\Ga(\pr\mnY(\mz), \pr),\nonumber\\
\mnY(\mz) &=& \con \{(1-\bc) +\bc [\sgY(\mz) +
\bb ] \},\nonumber
\\[-8pt]
\\[-8pt]
\sgY(\mz) &=&\sum_{j=1}^{\nPks} \kk(\mz;\pp_j,\ww_j)  \hh_j,\nonumber\\
\bb &=& \frac{\hh_0}{\ww_0}\exp\{{-(\mz-\ppa)/\ww_0}\} \one{\mz>\ppa}.
\nonumber
\end{eqnarray}
We discuss the approach we used to construct the distributions, employing
public knowledge where possible or default procedures otherwise. Parameter
values used for the different data sets are summarized in Table
\ref{tab:hyper}.

%t1 ###
\begin{table}
\tabcolsep=0pt
\caption{Prior hyperparameters for each of the real and simulated
data sets}
\label{tab:hyper}
\begin{tabular*}{\textwidth}{@{\extracolsep{\fill}}lcccccc@{}}
\hline
 & \multicolumn{6}{c@{}}{\textbf{Data set}} \\[-5pt]
\textbf{Prior}& \multicolumn{6}{c@{}}{\hrulefill}\\
\textbf{hyper-} &  &  &  & \multicolumn{1}{c}{\textbf{Single}} & \multicolumn{1}{c}{\textbf{Mean}} & \multicolumn{1}{c@{}}{\textbf{Simulation}} \\
\textbf{parameters} &\multicolumn{1}{c}{\textbf{Blank}} &\multicolumn{1}{c}{\textbf{Matrix}} &\multicolumn{1}{c}{\textbf{Known}} & \multicolumn{1}{c}{\textbf{lung}} & \multicolumn{1}{c}{\textbf{lung}} & \multicolumn{1}{c@{}}{\textbf{study}} \\\hline
% $l$ & 320 & 320 & 320 & 10 & 100 & 100 \\
%Range ( \kDa) & 2 - 75 & 2 - 75 & 2 - 75 & 2 - 75 & 2- 75 & 3+ \\
%Sampling Rate & 1:16 &1:16 & 1:16 &1:4 &1:4 &1:4 \\ \hline
%$J$ &
$\mJ$ & 20 & 20 & 20 & 100 & 100 & 150 \\
%$\hh_j$ &
$\gamR$ & 0.03 &0.03 & 0.03 &0.11 & 0.11 & 0.65 \\
$\eps$ & 0.79 & 0.79 & 0.79 &0.21 & 0.21 & 0.03 \\
%$\pp_j$ &
$T_0$ & 5.58 & 5.58 & 5.58 & 0.03 & 0.03 & 13.47 \\
$T_1$ & 217.14 & 217.14 &217.14 & 278.04 & 278.04 & 82.78 \\
%$\rr_j$ &
$\sigma^2_{\rho}$ & 0.1225 & 0.1225 & 0.1225 & 0.1225 & 0.1225 &
0.1225 \\
$\mu_\varrho$ &200 & 200 & 200 & 200 & 200 & 300 \\
$\sigma^2_\varrho$ & 0.49 & 0.49 & 0.49 & 0.49 & 0.49 & 0.49 \\
%$\pr$ &
$a_\pr$ & 0.25 & 0.25 & 0.25 & 0.25 & 0.25 & \multicolumn{1}{c@{}}{--} \\
$b_\pr$ & 0.02 & 0.11 & 25.91 & 1.14 & 0.17 & \multicolumn{1}{c@{}}{--} \\
%$\bc$ &
$a_\bc$ &1.36 & 8.33 &8.71 & 9.46 & 7.29 & 6.33 \\
$b_\bc$ &1 &1 &1 &1 &1 &1 \\
%$\hh_0$ &
%For iter=500k
%$\lambda_0$& 0.00009 & 0.0007 & 0.0015 & 0.0006 & 0.0004 & 0.0124 \\
%$\ww_0$ & $\hat{\omega}_0$ & 167.23 & 216.15 & 107.20 & 210.44 &
%268.05 & 11.22\\
%For iter=1mil
$\lambda_0$& 0.0009 & 0.0005 & 0.0012 & 0.0006 & 0.0004 & 0.0124 \\
%$\ww_0$ &
$\hat{\omega}_0$ & 171.68 & 290.14 & 127.30 & 210.96 & 268.52 & 11.22
\\
$\sigma^2_{\omega_0}$ & 0.25 & 0.25 & 0.25 & 0.25 & 0.25 & 0.25
\\ \hline
\end{tabular*}
\vspace*{-3pt}
\end{table}
%

%s4.1 ###
\subsection{Measurement error $\pr$ and overall level $\con$}\label{ss:con}

The exploratory data analysis of experimental spectra (see \Sec{ss:lh})
suggests that sample variances of $\{Y_{\mz}\}$ are nearly
proportional to
the mean.
%, with a mean-to-variance ratio $\phi$ under the gamma model.
We chose a gamma prior distribution $\pr\sim\Ga(a_\pr,b_\pr)$ for the
mean-to-variance ratio $\pr$, centered at a data-based value but
supporting a
wide range of prior uncertainty.
We binned the observations $\{Y_\mz\}$ of each data set in $50$ $\mu${s}-wide
blocks and calculated their block-specific means and variances. The prior
mean $a_\pr/b_\pr$ was set to the slope of a regression of the block
means on
block variances, with $a_\pr=0.25$ to attain a coefficient of
variation of
$2$.
% To construct a prior distribution for $\phi$, we regressed mean
%intensities
% on the variance of intensities in blocks of size 50 for each data set.
% Taking $\pr\sim\Ga(a_\pr,b_\pr)$ we equated the slope of the
%regression
% line with the prior mean, $a_\pr/b_\pr$, for $\phi$, with shape $a_
% to 0.25, or a coefficient of variation of $2$.

The parameter $\con$ may be interpreted as the mean level or scale for
$Y_\mz$, since $\E[ \sgY(\mz)]\approx1$ (see \Sec{ss:prosig}). Since
experimental levels depend on a wide range of exogenous variables and vary
widely among trials, it is difficult to elicit a subjective prior
distribution for this quantity. We instead employed a data-dependent
rescaling and set $\con\equiv\Yb$, the overall mean intensity. This is
comparable to rescaling the raw data by the average or total intensity.
Sensitivity analysis showed that this gave results very similar to those
under a tight data-dependent prior distribution.
%%%%%%%, \ie, $a_\con=100$ and $b_\con=100/\Yb$.

%s4.2 ###
\subsection{Prior distribution for protein signature
$\sgY(\cdot)$}\label{ss:prosig}

We specify a prior distribution for the protein signature
\[
\sgY(\mz) = \sum_{j=1}^{\nPks} \kk(\mz;\pp_j,\ww_j)  \hh_j
\]
by first specifying a distribution for $J$ and then, conditional on $J$,
taking $\{\pp_j,\ww_j, \hh_j\}$ to be i.i.d. from a specified joint distribution
[as in the infinitely divisible construction of the LARK models of
\citet
{WolpClydTu2006}]. To reflect uncertainty in the possible number of peaks,
\citeauthor{WolpClydTu2006} used a negative binomial distribution $\NB
(\aJ, \mJ)$ for~$J$ with mean and shape parameters $\mJ$ and $\aJ$. To
simplify elicitation while providing robust inference over a range of
spectra, we set $\aJ= 1$ throughout, leading to a geometric distribution:
\[
\P[J=j \mid\mJ] = %\binom{1+j-1}{j}
 \biggl(\frac{1}{1+\mJ} \biggr)
\biggl (\frac{\mJ}{1+\mJ} \biggr)^j,\qquad j\in\{0,1,\ldots \}
\]
with mean $\mJ$ representing the expected number of peaks for a given
experiment. \citet{CampWangHowaEtal2003} found approximately fifty
proteins for fractionated samples similar to the single and mean spectra
described in \Sec{ss:prot}, leading to perhaps seventy or so peaks due to
adducts, multiply charged ions, \etc. On this basis we chose $\mJ
=100$ with
a median of $\nPks\approx70$ peaks with symmetric 50\%, 90\% and 99\%
ranges of approximately $30\le\nPks\le140$, $5\le\nPks\le300$,
and $0.50
\le\nPks\le532.5$, respectively. For the blank, matrix and known protein
spectra, we set $\mJ= 20$.

There is little reason to give higher prior probability to one range of TOFs
than another without prior knowledge of the collection of proteins
present in
the samples. Thus, we take $\{\pp_j\}\jJ\iid\Un(\ppa,\ppb)$
(independently of
$\nPks$ and $\{\gamR_j\}\jJ$), for some interval of length $\dT=
\ppb-
\ppa$, large enough to include the TOF for all molecules of interest. To
eliminate saturation by matrix molecules at the low end, and to include as
wide as possible a range of the biologically relevant molecules, we
chose a
TOF interval corresponding to the range $[2~\mbox{kDa/e}\le\mbox{m$/$z}\le75~\mbox{kDa/e}
]$ for all
experimental data. Differing sampling rates and calibration levels for
different experiments lead the TOF ranges $[\ppa,\ppb]$ to vary across
experiments (see Table \ref{tab:hyper}).

We use expert opinion to construct an informed prior distribution on the
resolutions $\{\rho_j\}\jJ$ (see \Sec{ss:width}), which induces a
distribution
on the peak widths $\{\ww_j\}\jJ$. It has been suggested [\citet
{Siuz2003}, page~44] that individual peak resolutions $\rho_j$ \textit
{should} be nearly
constant across the entire TOF range, but in practice they are observed to
vary [see pages \mbox{6--32} of \citet{UsrGde-Voy}]. To
reflect this variation, we construct a~hierarchical prior distribution
for the
resolution parameters $\{\rho_j\}\jJ$ as follows. Independently of
the number
of peaks, TOFs and abundance, we take the resolutions to have log-normal
distributions,
\begin{eqnarray*}
\log( \RR) \mid\mnY_{\RR}, \sigma_\RR^2
&\sim&\N (\log(\mnY_{\RR}),\sigma_\RR^2 ), \\
\log(\rho_j) \mid\RR, \sigma_\rho^2
&\iid&\N (\log(\RR),\sigma_\rho^2 ),
\end{eqnarray*}
centered at an overall ``experiment'' resolution $\RR$ (which may be machine-
or condition-specific). We use the manufacturer's reported resolution ranges
[\citet{UsrGde-Voy}, Table 6-2 and Table H-6] for the Voyager
workstation and
set $\mnY_{\RR}= 200$ and set $\sigma^2_\RR= 0.49$, so  a
priori  the
distribution covers the range $50$--$800$ with $95\%$ probability, and
$32$--$928$ with $99\%$ probability. The standard deviation for the
individual resolutions was set to $\sigma_\rho=0.35$, seventy percent
of the
population standard deviation for resolution. For a~population
resolution of
$\mnY_{\RR} = 50$, this leads to a prior 99\% interval for individual
resolutions of $(20, 120)$, while at the upper extreme with a population
resolution of $\mnY_{\RR} = 800$, the prior 99\% interval covers the range
$(325, 1\mbox{,}710)$. Finally, the relationship between width, TOF and resolution
given by \Eqn{eqn:rr2ww} induces a log-normal prior distribution on
the width
parameters,
\[
\log(\ww_j) \mid\pp_j,\rho_j \ind\N
\bigl(\log(\pp_j/c\rho_j),\sigma_\rho^2 \bigr), \qquad j = 1, \ldots , J,
\]
with $c=2$ for the Cauchy kernel and $c=2\sqrt{\log4}$ for the Gaussian.

For protein abundances $\{\hh_j\}$ we use the left-truncated gamma
distribution $\Ga(0,\gamR,\eps)$ with parameters $\alpha=0$ and
$\gamR,
\eps$ (chosen below),
%. Here $\Ga(\alpha,\lambda,\eps)$ denotes the truncated gamma
%distribution
whose density function is given in general by
%
%e4.2 ###
\begin{equation}\label{eqn:tgam}
\Ga(\eta;\alpha,\lambda,\eps) \equiv
\frac{\lambda^\alpha}
{\Gamma(\alpha,\lambda\epsilon)}
\eta^{\alpha-1} e^{-\lambda\eta}\one{\eta>\eps},
\end{equation}
where $\Gamma(\alpha,x)\equiv\int_x^{\infty}z^{\alpha-1}e^{-z}\,
dz$ denotes
the complimentary incomplete gam\-ma function [\citet
{AbraSteg1964}, Section~6.5.3]. For $\alpha,\lambda>0$ this is the
conditional density for
a gamma-distributed $\Ga(\alpha,\lambda)$ random variable, given
that it
exceeds $\eps\ge0$; for strictly positive $\eps>0$, the distribution is
well-defined for all $\alpha\in\bbR$ including the limiting case
$\alpha=0$
[\citet{WolpClydTu2006}], which we adopt. The mean is
\[
\E[\hh] = \frac{1}{\gamR e^{\gamR\eps} \Eo(\gamR\eps)} ,
\]
where $\Eo(z)$ denotes the exponential integral function [\citet
{AbraSteg1964}, page 228]. The parameter $\eps$ may be interpreted as
the minimum
detectable abundance. In the limit $\eps\to0$, this distribution permits
an increasing number of isotopes with small abundance, while reflecting that
only a few isotopes are expected to have large abundance.

Discussions with spectrometrists suggest that the smallest peak that can
possibly be distinguished from noise is about 5--10\% of the average signal.
Using the midpoint of this interval, we take $\eps/\E[\hh]= \gamR
\eps
e^{\gamR\eps}  \Eo(\gamR\eps)=0.075$. For $t$ well away from the boundary
of $[\ppa,\ppb]$, $\int_\ppa^\ppb\kk(\mz;\pp_j,\ww_j)\,d\pp
_j\approx1$ (the
kernels are density functions), and
%
%e4.3 ###
\begin{equation}\label{eqn:Ef1}
\E [\sgY(t) ]
\approx\frac{\mJ} {\dT \gamR e^{\gamR\eps} \Eo(\gamR\eps)},
\qquad\ppa\ll t\ll\ppb.
\end{equation}
Because $f(t)$ is the normalized signal, we take $\E[\sgY(t)]=1$
 a
priori, leading to the solution $\eps=0.075\dT/\mJ$. Inverting the
function $\gamR\eps e^{\gamR\eps} \Eo(\gamR\eps)=0.075$, we
obtain $\gamR
\eps= 0.0227$, which determines $\gamR$ for any specified $\eps$. Values
for the different experiments are provided in \Tbl{tab:hyper}.

%s4.3 ###
\subsection{Prior distribution for matrix background}\label{ss:priMatrix}

As with other peaks, we use a log-normal distribution for the initial peak
width $\ww_0$ and a left-truncated gamma model for the abundance $\hh_0$.
Because the exponential decay varies greatly from experiment to experiment,
we utilize modestly informative priors based on the data. To construct
these, we first fit a linear regression with mean function
$\log (\bb )$ from equation (\ref{eqn:EK}) to the log
intensities in an
initial segment of the spectrum $(2~\mbox{kDa/e}<\mbox{m$/$z}<3.5~\mbox{kDa/e})$. We use the slope
and intercept from this fit to construct point estimates $\widehat{\ww_0}$
and $\widehat{\hh_0}$ to center the prior distributions
\[
\log(\ww_0) \sim\N (\log(\widehat{\ww_0}), \sigma^2_{\ww
_0} )
\]
with $\sigma^2_{\ww_0} = 0.25$ and
\[
\hh_0 \sim\Ga(0,\widehat{\gamR_0},\eps),
\]
where $\widehat{\gamR_0}$ is the solution to $ (\widehat{\gamR
_0}
e^{\eps\widehat{\gamR_0}}  \Eo(\eps\widehat{\gamR_0})
)^{-1} =
\widehat{\hh_0}$.

Finally, for the signal fraction $\bc$ we use a beta prior distribution
\[
\bc\sim\Be(a_\bc, b_\bc),
\]
with data-based mean $a_\bc/(a_\bc+b_\bc)= 1 - \Yb^{N}/\Yb$ and
%imaginary sample size
$b_\bc= 1$ (here $\Yb^N$ is the observed mean intensity in a ``noise''
region of the spectrum with low intensity and no apparent peaks, while
$\Yb$
is the overall mean intensity). With this choice the mode of the prior
density for $\bc$ is one whenever $\Yb> 2\Yb^N$, suggesting that the signal
dominates, and zero when the mean in the noise region exceeds half the
overall mean, favoring the thermal noise component over the nonparametric
signal model.

%s5 ###
\section{Posterior analysis}\label{s:POS}

To support inference about protein location and abundance, and about other
model parameters, we construct an ergodic Markov chain on the space
$\Theta$
of possible parameter vectors $\thb= \{\con, \nPks, \{\pp_j, \ww_j,
\break\hh_j\}\jJ, \bc, (\ww_0,  \hh_0), \lambda_J, \rho\}$ with the posterior
distribution as its stationary distribution. At each Markov chain step we
select one of the components of $\thb$ and either update it via a
Gibbs step
(replace the current value with a draw from its complete conditional
posterior distribution given the other components) or, if this is not
feasible, a random-walk Metropolis--Hastings (M--H) step by proposing a small
change in that component which is then accepted or rejected according
to the
Hastings probabilities. Note that each proposed change in $\nPks$
(which we
always take to be a random-walk step of size one) changes the
\textit{dimension} of $\thb$ (by three). Such dimension changing M--H
algorithms, introduced in \citet{Gree1995}, are called reversible jump MCMC
(RJ-MCMC) algorithms. Our approach is modeled after that of
\citet{WolpIcks2004}, where a general RJ-MCMC procedure for L\'{e}vy random
field models is presented. For updating the varying dimensional parameters
$\{\pp_j, \ww_j, \hh_j\}\jJ$ we consider the possible moves of
three basic
types: peak \textit{birth} [incrementing $\nPks$ by one and introducing
a new
triplet $(\pp_*,\ww_*, \hh_*)$]; peak \textit{death} [decrementing
$\nPks$ by
one and removing a randomly-chosen\vspace*{2pt} triplet $(\pp_j, \ww_j, \hh_j)$];
and peak
\textit{update} [moving a randomly-chosen triplet $(\pp_j, \ww_j, \hh_j)$
within $\bbR^3$]. We also incorporate two additional move types, peak
\textit{splitting}, in which a single peak is replaced by a pair of smaller
ones, and the reverse move, peak \textit{merging}, in which two nearby peaks
are replaced with a single larger one. These lead to a vast improvement in
algorithmic efficiency over RJ-MCMC algorithms using only birth/death and
update steps.

For sufficiently large spectra or complex protein mixtures, convergence to
the posterior distribution from random starting values may require
upward of
a million iterations. To reduce computation time, we begin the Markov chain
close to a mode, located using an EM algorithm [\citet{DempLairRubi1977}]
for a simple Gaussian approximation to our LARK model; see \citet{Hous2006}
for details of this and the RJ-MCMC algorithm.

%s5.1 ###
\subsection{Peak identification}\label{ss:PkID}

Features of the configuration $\{(\pp_j,\ww_j, \hh_j)\}\jJ$ are
updated at
each iteration of the RJ-MCMC sampler, with the number of peaks $\nPks
$ and
associated parameters changing. While the posterior mean of $\nPks$,
$\nPks\PM$, and the posterior mean function $\E[\mnY(\mz) \mid Y]$ [see
\Eqn{eqn:mod}] are well-defined quantities that may be used to
summarize the
RJ-MCMC output, the well known label switching problem complicates peak
identification. We have two ways of identifying peaks in an RJ-MCMC run.
The first is to use the $\nPks\HP$ peak locations $\{\tau_j\HP\}$
in the
single RC-MCMC iteration with highest posterior (HP) density, which is
proportional to the product of the likelihood function and prior density
evaluated at the parameter vector for that iteration. Alternatively, we may
use model averaging to identity local maxima in the denoised signal by
identifying the $\nPks\DV$ down-crossings of the derivative
$(d/dt)\E[\mu(t)\mid Y ]=\E[\mu'(t) \mid Y]$; these local modes
$\{\tau_j\DV\}$ are the collection of $\nPks\DV$ solutions of
%
%e5.1 ###
\begin{equation}\label{eqn:BMA}
\frac{d}{dt} \E[\mu(t-) \mid Y]>0>\frac{d}{dt} \E[\mu(t+) \mid Y].
\end{equation}
Because derivatives of densities may be used to construct wavelets, the
implied LARK model for the derivative process actually uses a continuous
wavelet dictionary based on the ``Mexican Hat'' (or Marr's) family
[\citet{Vida1999}, pages 48--49]. This method is closely related to the
recent paper of \citet{Nguyetal2010} who use zero-crossings of the
derivatives of Gaussian wavelets. Typically $\nPks\DV$ is smaller than
either $\nPks\PM$ or $\nPks\HP$ [one reason is that some pairs of
peaks with
small inter-peak distance $|\pp_j-\pp_k|$ will combine to generate a single
local maximum of $\mu(t)$, but all the major peaks will be represented among
the $\{\tau_j\DV\}$].

%s6 ###
\section{Examples}\label{s:imp}

In this section we illustrate our method using five data sets: a blank
spectrum, a matrix spectrum, a ``spiked'' spectrum from a~sample of known
protein composition, and two spectra (one single and one mean) for a serum
sample from a patient diagnosed with lung cancer. All data were generated
using the Voyager DE spectrometer [\citet{UsrGde-Voy}] at Duke University.
Prior hyperparameters were chosen as described in \Sec{s:PRI} and are given
in \Tbl{tab:hyper} for the five data sets. All examples in this
section use
the Cauchy kernel, selected because of its better fit to similar data
in a
preliminary investigation. %provided a better fit to segments of the
%spectrum
%with one isotopic peak.
% All RJ-MCMC
% were run for 250,000 iterations to ensure convergence (burn-in), with
% an additional 250,000 iterations used for posterior inference. {For
% the lung cancer data, which has more peaks, residual plots, traceplots
% and other diagnostics suggested that a longer run was necessary. We
% ran the chains for an additional 500,000 iterations. }
All RJ-MCMC were run for $500\mbox{,}000$ iterations to ensure convergence
(burn-in), with an additional $500\mbox{,}000$ iterations used for posterior
inference.

%f4 ###
\begin{figure}

\includegraphics{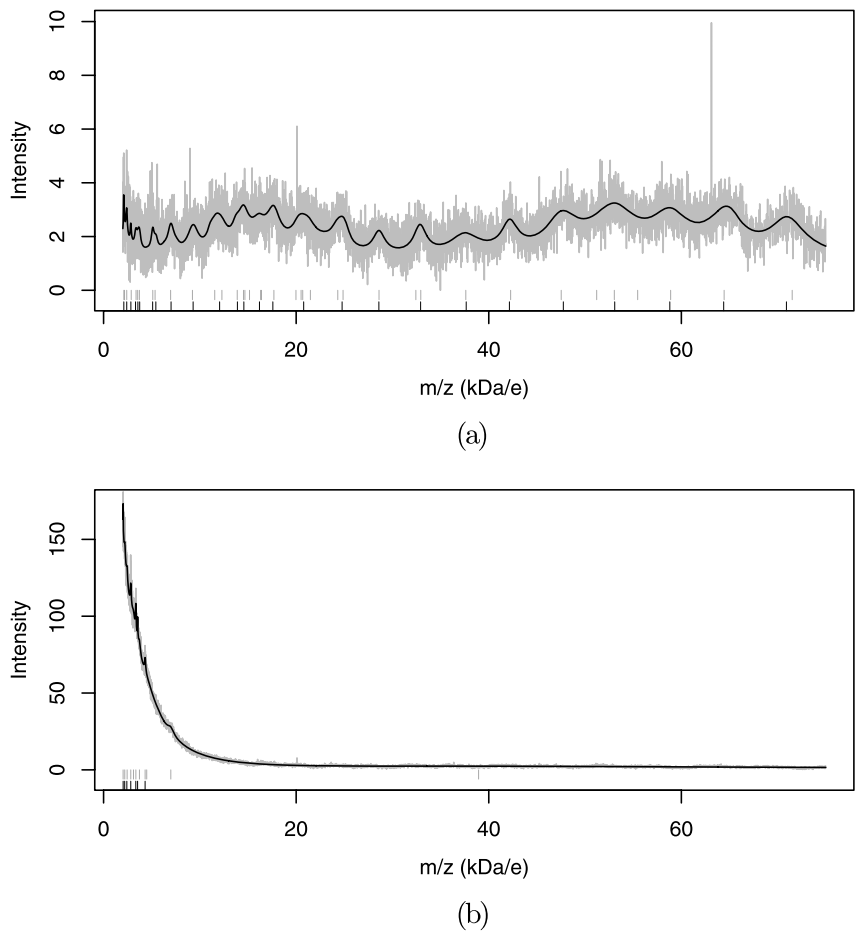}

\caption{The mean spectra based on ten
replicates for \textup{(a)} an empty plate illustrating thermal noise and (\textup{b}) the
sinapinic acid matrix with no proteins. The posterior means from LARK
are shown as solid curves with identified peak locations indicated as two
rows of tick marks: highest posterior realization (top, gray) and local
modes under model averaging (bottom, black). Note the difference in scale
of $Y$ axes.}\label{fig:blank} \label{fig:matrix}
\end{figure}

%s6.1 ###
\subsection{Blank spectrum}\label{ss:blank}

\Fig{fig:blank}(a) shows the recorded spectrum from the average of ten
blank-plate spectra each based on $32$ laser shots, with the posterior mean
$\E[\mu(t) \mid Y]$ shown as a solid curve. The two rows of tick
marks on
the horizontal axis represent peak locations identified using the highest
posterior realization $\{\tau_j\HP\}$ (top row) and local maxima
under model
averaging $\{\tau_j\DV\}$ (bottom). The highest-posterior realization
included $\nPks\HP=38$ peaks, the local modes under model averaging included
$\nPks\DV= 22$ peaks, while the posterior mean (and standard
deviation) were
$\nPks\PM=\E[\nPks\mid Y]=46.92$  $(4.22)$. With no solution or
matrix on
the metal plate, there can be no protein signature, but nevertheless the
spectrum shows numerous low-resolution peaks. The posterior expected
resolution in the blank spectrum was $\E[\RR\mid Y]=16.76$  $(2.92)$, lower
than the informative prior mean and significantly lower than the typical
resolutions for protein peaks in the other examples (see \Tbl{tbl:allRes1}).
These apparent peaks may reflect laser fluctuations or resonances in the
detector.
% Sinusoidal patterns appear at two frequencies, found by Fourier
%analysis to
% be approximately $10.66$ kHz and $91$ kHz (corresponding to periods of
% $94\ms $ and $11\ms$).
The number of periods and intensities may vary unpredictably across
spectroscopic samples, so rather than use a harmonic function as in
\citet
{HareWuetal2008}, we instead allow the adaptive LARK model to identify and
fit these low-resolution peaks as part of $\sgY( \cdot )$. They may be
discriminated from protein peaks in post-processing by their lower
resolution.

%t2 ###
\begin{table}
\tabcolsep=0pt
\caption{Posterior means and (standard deviations) for model parameters
for the five experimental data sets}\label{tbl:allRes1}
\begin{tabular*}{\textwidth}{@{\extracolsep{\fill}}lccd{2.3}d{3.3}d{4.3}d{2.3}cc@{}}
\hline
\textbf{Data set}
&\multicolumn{1}{c}{$\bolds\bc$}
&\multicolumn{1}{c}{$\bolds\pr$} &\multicolumn{1}{c}{$\bolds\RR$} &
\multicolumn{1}{c}{$\bolds{\hh_0}$}
&\multicolumn{1}{c}{$\bolds{\ww_0}$} &\multicolumn{1}{c}{$\bolds{\nPks\PM}$} &
\multicolumn{1}{c}{$\bolds{\nPks\HP}$} &$\bolds{\nPks\DV}$\\
\hline
Blank& 0.5660 & 7.385 & 16.76 & 38.59 & 233.70 & 46.92&38 &22\\
   & (0.0537) & (0.126) &
(2.92) & (11.14) & (150.80) & (4.22) &
  &  \\
 [3pt]
Matrix& 0.9569 & 6.770 & 15.03 & 163.70 & 17.41 & 12.27 & 11&\hphantom{5}7\\
   & (0.0063) & (0.121) &
(1.70) & (2.63) & (0.21) & (0.45) &
  &   \\
 [3pt]
Known& 0.9016 & 2.609 & 96.43 & 78.26 & 22.80 & 42.24 &42 &28 \\
   & (0.0021) & (0.093) &
(5.54) & (2.87) & (0.95) & (0.96) &
  &   \\
 [3pt]
Single& 0.9996 & 0.310 & 49.46 & 182.30 & 16.76 & 52.68 &45 &45 \\
   & (0.0004) & (0.005) &
(3.56) & (1.77) & (0.15) & (2.65) &
  &  \\
 [3pt]
Mean& 0.9846 & 3.952 & 69.35 & 188.40 & 15.77 & 76.91 &74 & 54\\
   & (0.0031) & (0.066) &
(3.55) & (0.97) & (0.07) & (1.47) &
  &   \\
\hline
\end{tabular*}
\end{table}

%s6.2 ###
\subsection{Matrix spectrum}\label{ss:matrix}

\Fig{fig:matrix}(b) shows the average of ten spectra, each based on $32$
laser shots from a sinapinic matrix solution containing no protein serum
sample. The posterior mean under the LARK model shows the characteristic
near-exponential spectral fall-off arising from the very low-molecular-weight
sinapinic acid matrix ions, as well as several low-resolution peaks
[posterior mean for $\RR= 15.03$  $(1.70)$] comparable to those in the blank
spectrum.

%s6.3 ###
\subsection{Known protein spectrum}\label{ss:known}

\begin{figure}

\includegraphics{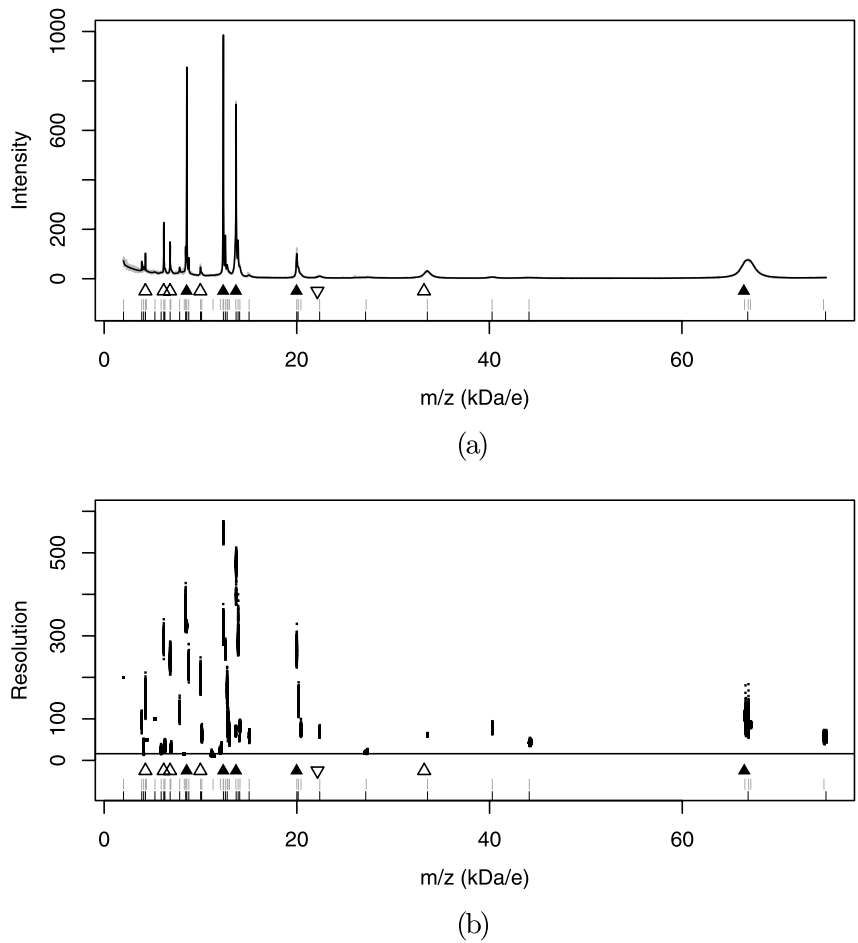}

\vspace*{-5pt}
\caption{LARK posterior mean (solid line) and data from a mixture of five
known proteins \textup{(a)} and associated posterior distribution of resolution
\textup{(b)}. Solid triangles represent singly-charged molecules, open triangles
represent doubly-charged molecules and the inverted triangle represents a
triply-charged protein. The rows of tick marks represent identified peak
locations using the highest posterior draw (top, gray) and local modes
under model averaging (bottom, black). The horizontal line in \textup{(b)}
corresponds to the median resolution from the blank spectrum
analysis.} \label{fig:known}
\vspace*{-3pt}
\end{figure}

\Fig{fig:known} shows the average of ten spectra (each with $32$
shots) from
a preparation of five proteins with known masses provided by Professor
M.\
Fitzgerald in the Department of Chemistry at Duke. The five known
masses of
singly-charged molecules are indicated by solid triangles and the five peaks
for doubly-charged molecules are indicated by open triangles;
% doubling the charge reduces the $\mtz$ ratio by half, so
each open triangle for a doubly-charged peak lies at one-half the m$/$z
value of the singly-charged peak for the same molecule. Finally, one
triply-charged peak at $22.1$~kDa/e (one-third the singly-charged
value) is
indicated by an inverted triangle. Peaks identified by our procedure are
indicated by vertical tick marks; these include all eleven ``true'' peaks,
plus several additional peaks. These may reflect contaminants, differential
isotopic compositions or thermal noise. Several of these identified peaks
have resolutions in the range of the median resolution for the blank spectrum
(\Tbl{tbl:allRes1}), suggesting that the model is capturing the
thermal noise
component. The peak at $3.903$~kDa (just below the smallest ``true'' peak)
has higher resolution than typical thermal peaks, and also higher abundance.
It is clearly present in all ten replicates, suggesting a potential
contaminant in the mixture. Features of the LARK model, such as the
resolution and abundance parameters, may aid in reducing false
positives and
prioritizing masses for further study, beyond using estimated mass alone.

%f5 ###

%s6.4 ###
\subsection{Lung cancer protein spectrum}\label{ss:prot}

\begin{figure}

\includegraphics{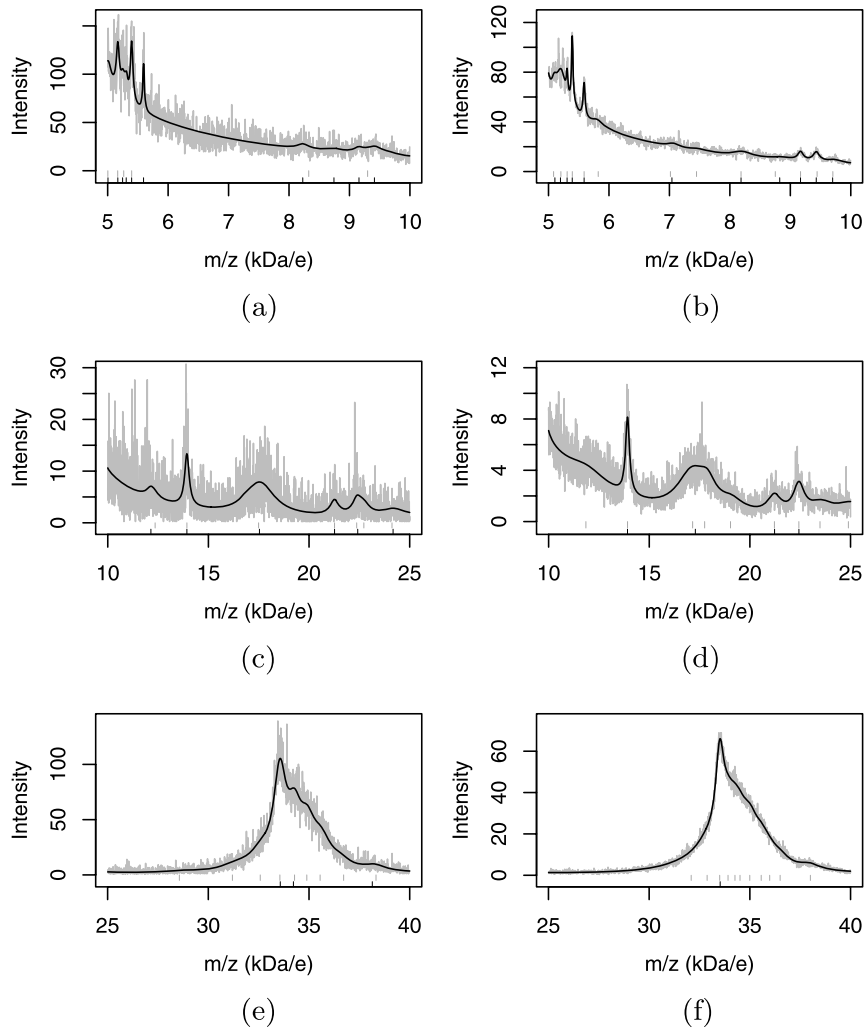}

\vspace*{-5pt}
\caption{Peak reconstruction using the LARK posterior mean (solid
line) for
the lung cancer patient using the single spectrum (left) and mean spectrum
(right) for segments of the respective spectra. The two rows of tick marks
on the horizontal axis correspond to identified peak locations using the
local maxima from highest probability draw (top, gray) and model averaging
(bottom, black).}
\label{fig:lung}
\vspace*{-5pt}
\end{figure}

\Fig{fig:lung} displays posterior reconstructions from LARK for different
segments of the single and mean spectra for the complete data depicted in
\Fig{fig:MS}. The noise reduction from averaging several spectra
results in
higher estimated precision $\pr$, just as one would expect
(\Tbl{tbl:allRes1}), approximately ten times higher than that for a~single
spectrum. The resolution is also higher in the mean spectrum,
% than the single spectrum,
leading to the identification of a larger number of peaks. Posterior means
for other fixed dimensional summaries are comparable for the single and mean
spectra. \Fig{fig:lung}(d)--(f) illustrates the ability of the LARK
model to
deconvolve a single large peak into several peaks. While there is a local
mode at $33.5$~kDa, corresponding to the doubly charged peak for albumin
(mass $67$~kDa), the relationship between resolution and mass
suggests that
there are other molecules present that lead to the wider than expected peak
and its asymmetric shape. The highest posterior realization from LARK
provides a~way to estimate these masses (in contrast, local modes of the
estimated signal would suggest just a single protein, while methods that
identify peaks by regions where the estimated signal is greater than some
specified threshold would report almost the entire range between
$31$--$39$~kDa).

%f6 ###

\setcounter{footnote}{4}

%s7 ###
\section{Simulation study}\label{s:sim}

Coombes et~al. (\citeyear{CoomKoometal2005}) construct a mathematical model for MALDI-TOF mass
spectrometry based on the physics of the process, providing a virtual mass
spectrometer. \citet{MorrCoometal2005} use this simulator to generate
spectra to explore operating characteristics of their mean-spectrum
undecimated wavelet threshold (MUDWT) peak detection method. They generated
$100$ data sets, each comprised of $100$ simulated spectra generated with
$150$ true peaks and additive i.i.d. Gaussian errors (sd $\sigma= 66$) [for
details, see Section~4.3 of \citet{MorrCoometal2005}; raw data and
code are
available on the \citet{Crom2004} website]. Although in theory the
simulated spectra should not need alignment before calculating mean spectra,
our exploratory analysis revealed boundary effects in the initial segment
(below approximately $3$~kDa) due to averaging, and visual peaks
outside of
the true peak mass $\pm0.3$\% tolerance window used by \citet
{MorrCoometal2005} to classify matches of identified peaks to true
peaks.\footnote{Jeffrey S. Morris  (M. D. Anderson), personal
communication.} Thus, for this
analysis we use mean spectra from each of the $100$ data sets for all masses
greater than $3$~kDa. In addition to the MUDWT approach of
\citeauthor
{MorrCoometal2005} and LARK, we identified peaks using the \texttt{R}
package \texttt{PROcess} [\citet{Li2005}],
which provides automatic baseline subtraction and peak identification.

Because the simulated data were generated using Gaussian errors with constant
variance, we replaced the gamma likelihood of the LARK model with a Gaussian
likelihood, with mean $\mnY(\mz)$ and constant precision $\phi=
1/\sigma^2$.
Otherwise, all model parameters have the same interpretation as before.
Preliminary investigation for a region of the spectra with a single known
peak indicated that the Gaussian kernel provided a better fit than the Cauchy
kernel we used for the experimental data. This also suggested that the
simulated spectra featured higher resolution than those of the Voyager
machine, leading us to select $\mnY_\RR=300$, keeping $\sigma_\RR=
0.49$ as
before. Following \citeauthor{MorrCoometal2005}, we use a robust estimate
of $\sigma$ from the wavelet decomposition as an alternative to the
data-based prior described in \Sec{s:PRI} for the nonconstant variance
model. For all remaining parameters, we used the default methods described
in \Sec{s:PRI} to specify hyperparameters (see \Tbl{tab:hyper}). We
ran the
RJ-MCMC algorithm for one million iterations, half for burn-in and half for
posterior inference. Peaks were identified using the local mode under model
averaging and the highest posterior realization.

% The results for the mean
% spectrum as follows:
%
%t3 ###
\begin{table}
\tabcolsep=0pt
\tablewidth=295pt
\caption{Summaries of true positive rates (TPR) and false discovery
rates (FDR) for $100$ simulated mean spectra using the mean-spectrum
undecimated wavelet transform (MUDWT) [Morris et~al. (\protect\citeyear{MorrCoometal2005})], LARK
with highest posterior realization (LARK-HP), LARK with local modes under
model averaging (LARK-MA)~and~PROcess [Li (\protect\citeyear{Li2005})]} \label{tab:sim}
\begin{tabular*}{295pt}{@{\extracolsep{\fill}}lcd{1.4}c@{}}
\hline
\textbf{Summary} & \textbf{Method} &\multicolumn{1}{c}{\textbf{Average}}&\multicolumn{1}{c@{}}{\textbf{95\% range}}\\
\hline
TPR & MUDWT & 0.83 & 0.77--0.89 \\
& LARK-HP & 0.82 & 0.75--0.89 \\
& LARK-MA & 0.73 & 0.61--0.81 \\
& PROcess & 0.31 & 0.10--0.43 \\ [4pt]
FDR & MUDWT & 0.06 & 0.003--0.208 \\
& LARK-HP & 0.05 & 0.000--0.132 \\
& LARK-MA & 0.01 & 0.000--0.050 \\
& PROcess & 0.0033 & 0.000--0.032 \\ \hline
\end{tabular*}
\end{table}

%With more peaks in the simulated data than in our previous fractionated
%experiments,
For all methods, true peaks were classified as a \textit{true positive} or
\textit{true discovery} if the mass of any identified peak was within
$\pm
0.3\%$ of a true mass [as in \citet{MorrCoometal2005}]. Identified
peaks outside of the $\pm0.3\%$ tolerance window of true peaks were regarded
as \textit{false positives}. The True Positive Rate (TPR), the
proportion of
true discoveries, and False Discovery Rate (FDR), the proportion of false
positives out of all identified peaks, were calculated for each of the 100
simulated mean spectra and are summarized in \Tbl{tab:sim}. Overall,
\texttt{PROcess} has the lowest average FDR of all methods---however, its
TPR is much worse than either of the LARK or MUDWT methods, which estimate
the baseline and denoise simultaneously. Peak identification using
down-crossings under model averaging (LARK-MA) has the better FDR of
the two
LARK methods, but has a lower TPR because local modes may miss overlapping
peaks. Peak identification using the LARK HP realization and the wavelet
method provide comparable TPR and FDR performance, with the MUDWT method
having a slightly better TPR, while the LARK-HP method has a slightly
improved FDR. The absolute differences of TPR and FDR rates for the two
methods are both about $0.01$, ``statistically significant'' using a paired
$t$-test but not practically significant, leading to the discovery of an
extra 1--2 proteins by MUDWT (on average) at the cost of a~comparable number
of additional false positives. A further breakdown of the TPR for LARK by
prevalence and abundance [as in \citet{MorrCoometal2005}, Table 4]
is provided in the supplemental materials [\citet
{peaks-supp2010}], with LARK-HP having substantially higher TPR than
MUDWT for peaks in the higher prevalence groups across all abundance
categories, but poorer performance than MUDWT for the two lowest
prevalence groups.

%s8 ###
\section{Discussion}\label{s:dis}

The Gaussian LARK model leads to true positive rates and false discovery
rates comparable to adaptive nonparametric wavelet methods for simulated
Gaussian intensities data, while the gamma LARK model is able to
capture the
mean/variance relationship that is observed in experimental data.
Exploratory data analysis may be used to decide which model is more
appropriate by examining the mean/variance relationship or residual analysis,
with other error models easily substituted to define alternative likelihood
functions given the mean function.

A key feature of the LARK methodology is the ability to deconvolve large
peaks into mixtures of protein signatures. The model is able to
identify all
masses for the laboratory experiment with known protein mixture, even though
many of the doubly charged proteins have low abundance. To capture the
multiply charged nature of proteins in MALDI-TOF even more effectively, LARK
models may be constructed using a kernel tailored to this purpose as a
mixture of two peaks centered at the singly and doubly charged masses,
with a
mixing weight to control the relative abundance of the two charges. This
constraint reduces the number of free parameters and may lead to a more
efficient algorithm. The method described in this paper is intended for use
with either a single spectrum or a mean spectrum. Hierarchical versions of
the LARK model for MALDI-TOF data are under development for modeling multiple
spectra, which provide automatic calibration of multiple spectra and permit
classification of subjects into groups.

The LARK models are implemented in an \texttt{R} [\citet{R2010}]
package, with a shared library written in \texttt{C} and
\texttt{FORTRAN} for the RJ-MCMC algorithm. Although the LARK model
for peak identification is more computationally intensive than the
wavelet method of \citet{MorrCoometal2005}, with 10,000 iterations
taking 10 minutes on a dual $3$~GHz Quad Core Xeon Mac Pro for the
simulation study (running on a single processor), its running times
increase only linearly with the number of peaks and volume of data,
since no matrix inversion is required. Despite the computational
overhead of RJ-MCMC, \citet{ClydWolp2007} and \citet{WolpClydTu2006} have
demonstrated that LARK models can provide significant reductions in
mean squared error in comparison with some of the best wavelet methods
such as the nondecimated wavelet approach of \citet{JohnSilv2005a}
and the continuous wavelets of \citet{ChuClydLian2009}. Future
work will incorporate advances in adaptive MCMC methods which may
accelerate convergence for random-walk update steps or lead to
improved proposal distributions for peak birth based on residuals or
peak death utilizing abundance. The software is available from the
first author's website.

\section*{Acknowledgments}
The authors would like to thank Michael J.~Campa,
Michael C.~Fitzgerald, Nayela Khan, Sheila Lee, Edward Patz, Jr. and
Petra~L.~Roulhac at Duke University for providing the
experimental data used in our examples and Jeff Morris from M.D.
Anderson for providing the simulated data. We have benefited
greatly from many fruitful discussions with these individuals. We
also wish to thank the anonymous referees, Associate Editors and the
Editor for their helpful comments on earlier versions of this
manuscript.
% to make the software available from CRAN in the near future.

% Incorporating information about resolution and peak shape leads to
%improved
% peak detection. Protein-protein interactions, multiple isotopes, and
%other
% factors may lead to multiple apparent peaks for a single protein. An
% experimenter familiar with the molecular masses and chemical
%properties of
% common molecular species may be able to resolve the cluster of peaks
% representing a single protein and the sum the components to find the
%total
% protein abundance.

%The single spectrum model developed in this paper provides the first
%stage
%prior distribution for hierarchical models for multiple (exchangeable)
%spectra \citep{Hous2006}.
%
%The analysis of multiple spectra is complicated by misalignment due to
%the
%variability of TOFs across shots, even within the same subject.
% or experimental condition.
%
%Simply averaging misaligned spectra across shots %for the same subject
%leads to additional difficulties, such as the broadening of peaks or
%even the
%loss of small peaks.
%
%Automatic calibration and alignment of spectra in this hierarchical
%approach
%is achieved by centering the individual TOF parameters around those
%from the
%first stage. TOFs may vary across spectra, but they remain centered at
%subject-specific expected TOFs, leading to properly aligned TOFs.
%
%This hierarchical formulation accommodates subjects within subgroups
%(cases
%and controls, for example) and uses additional ``mark'' parameters to
%distinguish which proteins are common across groups and which serve to
%differentiate the groups.

\begin{supplement}%[id=suppA]
\stitle{Additional results for the simulation study\\}
\slink[doi,text={10.1214/10-\break
AOAS450SUPP}]{10.1214/10-AOAS450SUPP} %[doi,text={...}] - jei reikia
%suskaldyti doi
\slink[url]{http://lib.stat.cmu.edu/aoas/450/supplement.pdf}
\sdatatype{.pdf}
\sdescription{True positive rates for LARK estimates from the
simulation study broken down by peak prevalence and
average intensity of peaks across samples.}
\end{supplement}

% imsref loaded by smiklovaite, 2011-03-28 08:33:19
% imsref loaded by smiklovaite, 2011-03-28 10:00:50
%

\printaddresses


\begin{thebibliography}{41}
% BibTex style file: ims.bst, 2010-03-23
% Default style options (sort=0,type=number).
% Used options (sort=1,type=nameyear).

%b1 ###
\bibitem[\protect\citeauthoryear{Abramowitz and Stegun}{1964}]{AbraSteg1964}
%
\begin{bbook}[mr]
\bauthor{\bsnm{Abramowitz},~\bfnm{Milton}\binits{M.}} \AND
\bauthor{\bsnm{Stegun},~\bfnm{Irene~A.}\binits{I.~A.}}, eds.
(\byear{1964}).
\btitle{Handbook of Mathematical Functions with Formulas, Graphs, and
Mathematical Tables}.
\bseries{National Bureau of Standards Applied Mathematics Series}
\bvolume{55}.
\bpublisher{For sale by the Superintendent of Documents, U.S. Government
Printing Office}, \baddress{Washington, DC}.
\bid{mr={0167642}}
\end{bbook}
%
\endbibitem

%b2 ###
\bibitem[\protect\citeauthoryear{Applied Biosystems}{2001}]{UsrGde-Voy}
%
\begin{bmisc}[auto:STB|2011-03-03|12:04:44]
\borganization{Applied Biosystems}
(\byear{2001}).
\bhowpublished{\textit{Voyager Biospectrometry Workstation with Delayed
Extraction Technology User Guide Version 5.1}. Applied Biosystems, Foster
City, CA}.
\end{bmisc}
%
\endbibitem

%b3 ###
\bibitem[\protect\citeauthoryear{Baggerly, Coombes and
Morris}{2006}]{BaggCoomMorr2006}
%
\begin{bincollection}[auto:STB|2011-03-03|12:04:44]
\bauthor{\bsnm{Baggerly},~\bfnm{K.~A.}\binits{K.~A.}},
\bauthor{\bsnm{Coombes},~\bfnm{K.~R.}\binits{K.~R.}} \AND
\bauthor{\bsnm{Morris},~\bfnm{J.~S.}\binits{J.~S.}}
(\byear{2006}).
\btitle{An introduction to high-throughput bioinformatics data}.
In \bbooktitle{Bayesian Inference for Gene Expression and Proteomics}
(\beditor{\bfnm{Kim-Ahn}\binits{K.-A.}~\bsnm{Do}},
\beditor{\bfnm{Peter}\binits{P.}~\bsnm{M{\"u}ller}} \AND
\beditor{\bfnm{Marina}\binits{M.}~\bsnm{Vannucci}}, eds.)
\bpages{Chapter~1, 1--39}.
\bpublisher{Cambridge Univ. Press}, \baddress{Cambridge}.
\end{bincollection}
%
\endbibitem

%b4 ###
\bibitem[\protect\citeauthoryear{Baggerly, Morris and
Coombes}{2004}]{BaggMorrCoom2004}
%
\begin{barticle}[pbm]
\bauthor{\bsnm{Baggerly},~\bfnm{Keith~A.}\binits{K.~A.}},
\bauthor{\bsnm{Morris},~\bfnm{Jeffrey~S.}\binits{J.~S.}} \AND
\bauthor{\bsnm{Coombes},~\bfnm{Kevin~R.}\binits{K.~R.}}
(\byear{2004}).
\btitle{Reproducibility of SELDI-TOF protein patterns in serum: Comparing
datasets from different experiments}.
\bjournal{Bioinformatics}
\bvolume{20}
\bpages{777--785}.
\bid{doi={10.1093/bioinformatics/btg484}, issn={1367-4803}, pii={btg484},
pmid={14751995}}
\end{barticle}
%
\endbibitem

%b5 ###
\bibitem[\protect\citeauthoryear{Campa et~al.}{2003}]{CampWangHowaEtal2003}
%
\begin{barticle}[auto:STB|2011-03-03|12:04:44]
\bauthor{\bsnm{Campa},~\bfnm{M.~J.}\binits{M.~J.}},
\bauthor{\bsnm{Wang},~\bfnm{M.~Z.}\binits{M.~Z.}},
\bauthor{\bsnm{Howard},~\bfnm{B.~A.}\binits{B.~A.}},
\bauthor{\bsnm{Fitzgerald},~\bfnm{M.~C.}\binits{M.~C.}} \AND
\bauthor{\bsnm{Patz},~\bfnm{E.~F.}\binits{E.~F.}~\bsuffix{Jr.}}
(\byear{2003}).
\btitle{Protein expression profiling identifies MIF and Cyclophilin~A as
potential molecular targets in non-small cell lung cancer}.
\bjournal{Cancer Research}
\bvolume{63}
\bpages{1652--1656}.
\end{barticle}
%
\endbibitem

%b6 ###
\bibitem[\protect\citeauthoryear{Chu, Clyde and
Liang}{2009}]{ChuClydLian2009}
%
\begin{barticle}[mr]
\bauthor{\bsnm{Chu},~\bfnm{Jen-Hwa}\binits{J.-H.}},
\bauthor{\bsnm{Clyde},~\bfnm{Merlise~A.}\binits{M.~A.}} \AND
\bauthor{\bsnm{Liang},~\bfnm{Feng}\binits{F.}}
(\byear{2009}).
\btitle{Bayesian function estimation using continuous wavelet dictionaries}.
\bjournal{Statist. Sinica}
\bvolume{19}
\bpages{1419--1438}.
\bid{issn={1017-0405}, mr={2589190}}
\end{barticle}
%
\endbibitem

%b7 ###
\bibitem[\protect\citeauthoryear{Clyde, House and Wolpert}{2006}]{ClydHousWolp2006}
%
\begin{bincollection}[auto:STB|2011-03-03|12:04:44]
\bauthor{\bsnm{Clyde},~\bfnm{M.~A.}\binits{M.~A.}},
\bauthor{\bsnm{House},~\bfnm{L.~L.}\binits{L.~L.}} \AND
\bauthor{\bsnm{Wolpert},~\bfnm{R.~L.}\binits{R.~L.}}
(\byear{2006}).
\btitle{Nonparametric models for proteomic peak identification and
quantification}.
In \bbooktitle{Bayesian Inference for Gene Expression and Proteomics}
(\beditor{\bfnm{Kim-Ahn}\binits{K.-A.}~\bsnm{Do}},
\beditor{\bfnm{Peter}\binits{P.}~\bsnm{M{\"u}ller}} \AND
\beditor{\bfnm{Marina}\binits{M.}~\bsnm{Vannucci}}, eds.)
\bpages{Chapter~15, 293--308}.
\bpublisher{Cambridge Univ. Press}, \baddress{Cambridge}.
\end{bincollection}
%
\endbibitem

%b8 ###
\bibitem[\protect\citeauthoryear{Clyde and Wolpert}{2007}]{ClydWolp2007}
%
\begin{bincollection}[mr]
\bauthor{\bsnm{Clyde},~\bfnm{Merlise~A.}\binits{M.~A.}} \AND
\bauthor{\bsnm{Wolpert},~\bfnm{Robert~L.}\binits{R.~L.}}
(\byear{2007}).
\btitle{Nonparametric function estimation using overcomplete dictionaries}.
In \bbooktitle{Bayesian Statistics 8}
(\beditor{\bfnm{J.~M.}\binits{J.~M.}~\bsnm{Bernardo}},
\beditor{\bfnm{M.~J.}\binits{M.~J.}~\bsnm{Bayarri}},
\beditor{\bfnm{J.~O.}\binits{J.~O.}~\bsnm{Berger}},
\beditor{\bfnm{A.~P.}\binits{A.~P.}~\bsnm{Dawid}},
\beditor{\bfnm{D.}\binits{D.}~\bsnm{Heckerman}},
\beditor{\bfnm{A.~F.~M.}\binits{A.~F.~M.}~\bsnm{Smith}} \AND
\beditor{\bfnm{M.}\binits{M.}~\bsnm{West}}, eds.)
\bpages{91--114}.
\bpublisher{Oxford Univ. Press}, \baddress{Oxford}.
\bid{mr={2433190}}
\end{bincollection}
%
\endbibitem

%b9 ###
\bibitem[\protect\citeauthoryear{Coombes et~al.}{2005a}]{CoomKoometal2005}
%
\begin{barticle}[auto:STB|2011-03-03|12:04:44]
\bauthor{\bsnm{Coombes},~\bfnm{K.~R.}\binits{K.~R.}},
\bauthor{\bsnm{Koomen},~\bfnm{J.~M.}\binits{J.~M.}},
\bauthor{\bsnm{Baggerly},~\bfnm{K.~A.}\binits{K.~A.}},
\bauthor{\bsnm{Morris},~\bfnm{J.~S.}\binits{J.~S.}} \AND
\bauthor{\bsnm{Kobayashi},~\bfnm{R.}\binits{R.}}
(\byear{2005}a).
\btitle{Understanding the characteristics of mass spectrometry data
through the
use of simulation}.
\bjournal{Cancer Informatics}
\bvolume{1}
\bpages{41--52}.
\end{barticle}
%
\endbibitem

%b10 ###
\bibitem[\protect\citeauthoryear{Coombes et~al.}{2005b}]{CoomTsavetal2005}
%
\begin{barticle}[auto:STB|2011-03-03|12:04:44]
\bauthor{\bsnm{Coombes},~\bfnm{K.~R.}\binits{K.~R.}},
\bauthor{\bsnm{Tsavachidis},~\bfnm{S.}\binits{S.}},
\bauthor{\bsnm{Morris},~\bfnm{J.~S.}\binits{J.~S.}},
\bauthor{\bsnm{Baggerly},~\bfnm{K.~A.}\binits{K.~A.}},
\bauthor{\bsnm{Hung},~\bfnm{M.~C.}\binits{M.~C.}} \AND
\bauthor{\bsnm{Kuerer},~\bfnm{H.~M.}\binits{H.~M.}}
(\byear{2005}b).
\btitle{Improved peak detection and quantification of mass
spectrometry data
acquired from surface-enhanced laser desorption and ionization by denoising
spectra with the undecimated discrete wavelet transform}.
\bjournal{Proteomics}
\bvolume{5}
\bpages{4107--4117}.
\end{barticle}
%
\endbibitem

%b11 ###
\bibitem[\protect\citeauthoryear{Cromwell}{2004}]{Crom2004}
%
\begin{bmisc}[auto:STB|2011-03-03|12:04:44]
\borganization{Cromwell}
(\byear{2004}).
\bhowpublished{Cromwell Mat{L}ab package. M. D. Anderson Cancer Center,
Houston, TX. Available at
\url{http://bioinformatics.mdanderson.org/cromwell.html}}.
\end{bmisc}
%
\endbibitem

%b12 ###
\bibitem[\protect\citeauthoryear{Dass}{2001}]{Dass2001}
%
\begin{bbook}[auto:STB|2011-03-03|12:04:44]
\bauthor{\bsnm{Dass},~\bfnm{C.}\binits{C.}}
(\byear{2001}).
\btitle{Principles and Practice of Biological Mass Spectrometry}.
\bpublisher{Wiley},
\baddress{New York}.
\end{bbook}
%
\endbibitem

%b13 ###
\bibitem[\protect\citeauthoryear{Dempster, Laird and
Rubin}{1977}]{DempLairRubi1977}
%
\begin{barticle}[mr]
\bauthor{\bsnm{Dempster},~\bfnm{A.~P.}\binits{A.~P.}},
\bauthor{\bsnm{Laird},~\bfnm{N.~M.}\binits{N.~M.}} \AND
\bauthor{\bsnm{Rubin},~\bfnm{D.~B.}\binits{D.~B.}}
(\byear{1977}).
\btitle{Maximum likelihood from incomplete data via the {EM} algorithm}.
\bjournal{J. Roy. Statist. Soc. Ser. B}
\bvolume{39}
\bpages{1--38}.
\bnote{With discussion}.
\bid{issn={0035-9246}, mr={0501537}}
\end{barticle}
%
\endbibitem

%%b14 ###
%Vannucci}{2006}]{DoMullVann2006}
%%
%(\byear{2006}).
%%

%b15 ###
\bibitem[\protect\citeauthoryear{Franzen}{1997}]{Fran1997}
%
\begin{barticle}[auto:STB|2011-03-03|12:04:44]
\bauthor{\bsnm{Franzen},~\bfnm{J.}\binits{J.}}
(\byear{1997}).
\btitle{Improved resolution for {MALDI-TOF} mass spectrometers: A mathematical
study}.
\bjournal{International Journal of Mass Spectrometry and Ion Processes}
\bvolume{164}
\bpages{19--34}.
\end{barticle}
%
\endbibitem

%b16 ###
\bibitem[\protect\citeauthoryear{Green}{1995}]{Gree1995}
%
\begin{barticle}[mr]
\bauthor{\bsnm{Green},~\bfnm{Peter~J.}\binits{P.~J.}}
(\byear{1995}).
\btitle{Reversible jump {M}arkov chain {M}onte {C}arlo computation and
{B}ayesian model determination}.
\bjournal{Biometrika}
\bvolume{82}
\bpages{711--732}.
\bid{doi={10.1093/biomet/82.4.711}, issn={0006-3444}, mr={1380810}}
\end{barticle}
%
\endbibitem

%b17 ###
\bibitem[\protect\citeauthoryear{Guindani et~al.}{2006}]{GuinDoetal2006}
%
\begin{bincollection}[auto:STB|2011-03-03|12:04:44]
\bauthor{\bsnm{Guindani},~\bfnm{M.}\binits{M.}},
\bauthor{\bsnm{Do},~\bfnm{K.~A.}\binits{K.~A.}},
\bauthor{\bsnm{M{\"{u}}ller},~\bfnm{P.}\binits{P.}} \AND
\bauthor{\bsnm{Morris},~\bfnm{J.~S.}\binits{J.~S.}}
(\byear{2006}).
\btitle{Bayesian mixture models for gene expression and protein profiles}.
(\beditor{\bfnm{Kim-Ahn}\binits{K.-A.}~\bsnm{Do}},
\beditor{\bfnm{Peter}\binits{P.}~\bsnm{M{\"u}ller}} \AND
\beditor{\bfnm{Marina}\binits{M.}~\bsnm{Vannucci}}, eds.)
\bpages{Chapter~12, 238--253}.
\bpublisher{Cambridge Univ. Press}, \baddress{Cambridge}.
\end{bincollection}
%
\endbibitem

%b18 ###
\bibitem[\protect\citeauthoryear{Harezlak et~al.}{2008}]{HareWuetal2008}
%
\begin{barticle}[auto:STB|2011-03-03|12:04:44]
\bauthor{\bsnm{Harezlak},~\bfnm{J.}\binits{J.}},
\bauthor{\bsnm{Wu},~\bfnm{M.}\binits{M.}},
\bauthor{\bsnm{Wang},~\bfnm{M.}\binits{M.}},
\bauthor{\bsnm{Schwartzman},~\bfnm{A.}\binits{A.}},
\bauthor{\bsnm{Christian},~\bfnm{D.}\binits{D.}} \AND
\bauthor{\bsnm{Lin},~\bfnm{X.}\binits{X.}}
(\byear{2008}).
\btitle{Biomarker discovery for Arsenic exposure using functional data analysis
and feature learning of mass spectrometry proteomic data}.
\bjournal{Journal of Proteome Research}
\bvolume{7}
\bpages{217--224}.
\end{barticle}
%
\endbibitem

%b19 ###
\bibitem[\protect\citeauthoryear{House}{2006}]{Hous2006}
%
\begin{bmisc}[auto:STB|2011-03-03|12:04:44]
\bauthor{\bsnm{House},~\bfnm{L.~L.}\binits{L.~L.}}
(\byear{2006}).
\bhowpublished{Nonparametric Bayesian models in expression proteomic
applications. {Ph.D.} dissertation. Dept. Statist. Sci., Duke Univ., Durham,
NC}.
\end{bmisc}
%
\endbibitem

%b20 ###
\bibitem[\protect\citeauthoryear{House, Clyde and
Wolpert}{2011}]{peaks-supp2010}
%
\begin{bmisc}[auto:STB|2011-03-03|12:04:44]
\bauthor{\bsnm{House},~\bfnm{L.~L.}\binits{L.~L.}},
\bauthor{\bsnm{Clyde},~\bfnm{M.~A.}\binits{M.~A.}} \AND
\bauthor{\bsnm{Wolpert},~\bfnm{R.~L.}\binits{R.~L.}}
(\byear{2011}).
\bhowpublished{Supplement to ``Bayesian nonparametric models for peak
identification in MALDI-TOF mass spectroscopy.''
\href{http://dx.doi.org/10.1214/10-AOAS450SUPP}{DOI:10.1214/10-AOAS450SUPP}}.
\end{bmisc}
%
\endbibitem

%b21 ###
\bibitem[\protect\citeauthoryear{Johnstone and
Silverman}{2005}]{JohnSilv2005a}
%
\begin{barticle}[mr]
\bauthor{\bsnm{Johnstone},~\bfnm{Iain~M.}\binits{I.~M.}} \AND
\bauthor{\bsnm{Silverman},~\bfnm{Bernard~W.}\binits{B.~W.}}
(\byear{2005}).
\btitle{Empirical {B}ayes selection of wavelet thresholds}.
\bjournal{Ann. Statist.}
\bvolume{33}
\bpages{1700--1752}.
\bid{doi={10.1214/009053605000000345}, issn={0090-5364}, mr={2166560}}
\end{barticle}
%
\endbibitem\vadjust{\goodbreak}

%b22 ###
\bibitem[\protect\citeauthoryear{Kempka et~al.}{2004}]{Kempetal2004}
%
\begin{barticle}[auto:STB|2011-03-03|12:04:44]
\bauthor{\bsnm{Kempka},~\bfnm{M.}\binits{M.}},
\bauthor{\bsnm{S{\"o}dahl},~\bfnm{J.}\binits{J.}},
\bauthor{\bsnm{Bj{\"o}rk},~\bfnm{A.}\binits{A.}} \AND
\bauthor{\bsnm{Roeraade},~\bfnm{J.}\binits{J.}}
(\byear{2004}).
\btitle{Improved method for peak picking in matrix-assisted laser
desorption/ionization time-of-flight mass spectrometry}.
\bjournal{Rapid Communications in Mass Spectrometry}
\bvolume{18}
\bpages{1208--1212}.
\end{barticle}
%
\endbibitem

%b23 ###
\bibitem[\protect\citeauthoryear{Li}{2005}]{Li2005}
%
\begin{bmisc}[auto:STB|2011-03-03|12:04:44]
\bauthor{\bsnm{Li},~\bfnm{X.}\binits{X.}}
(\byear{2005}).
\bhowpublished{{PRO}cess: Ciphergen {SELDI-TOF} Processing.
\texttt{R}
Package Version 1.24.0. Available at
\href{http://www.bioconductor.org/help/bioc-views/2.6/bioc/html/PROcess.html}{http://www.bioconductor.org/help/bioc-views/2.6/bioc/html/}
\href{http://www.bioconductor.org/help/bioc-views/2.6/bioc/html/PROcess.html}{PROcess.html}}.
\end{bmisc}
%
\endbibitem

%b24 ###
\bibitem[\protect\citeauthoryear{Malyarenko et~al.}{2005}]{Malyetal2005}
%
\begin{barticle}[pbm]
\bauthor{\bsnm{Malyarenko},~\bfnm{Dariya~I.}\binits{D.~I.}},
\bauthor{\bsnm{Cooke},~\bfnm{William~E.}\binits{W.~E.}},
\bauthor{\bsnm{Adam},~\bfnm{Bao-Ling}\binits{B.-L.}},
\bauthor{\bsnm{Malik},~\bfnm{Gunjan}\binits{G.}},
\bauthor{\bsnm{Chen},~\bfnm{Haijian}\binits{H.}},
\bauthor{\bsnm{Tracy},~\bfnm{Eugene~R.}\binits{E.~R.}},
\bauthor{\bsnm{Trosset},~\bfnm{Michael~W.}\binits{M.~W.}},
\bauthor{\bsnm{Sasinowski},~\bfnm{Maciek}\binits{M.}},
\bauthor{\bsnm{Semmes},~\bfnm{O.~John}\binits{O.~J.}} \AND
\bauthor{\bsnm{Manos},~\bfnm{Dennis~M.}\binits{D.~M.}}
(\byear{2005}).
\btitle{Enhancement of sensitivity and resolution of surface-enhanced laser
desorption/ionization time-of-flight mass spectrometric records for serum
peptides using time-series analysis techniques}.
\bjournal{Clin. Chem.}
\bvolume{51}
\bpages{65--74}.
\bid{doi={10.1373/clinchem.2004.037283}, issn={0009-9147},
pii={clinchem.2004.037283}, pmid={15550476}}
\end{barticle}
%
\endbibitem

%b25 ###
\bibitem[\protect\citeauthoryear{Morris et~al.}{2005}]{MorrCoometal2005}
%
\begin{barticle}[pbm]
\bauthor{\bsnm{Morris},~\bfnm{Jeffrey~S.}\binits{J.~S.}},
\bauthor{\bsnm{Coombes},~\bfnm{Kevin~R.}\binits{K.~R.}},
\bauthor{\bsnm{Koomen},~\bfnm{John}\binits{J.}},
\bauthor{\bsnm{Baggerly},~\bfnm{Keith~A.}\binits{K.~A.}} \AND
\bauthor{\bsnm{Kobayashi},~\bfnm{Ryuji}\binits{R.}}
(\byear{2005}).
\btitle{Feature extraction and quantification for mass spectrometry in
biomedical applications using the mean spectrum}.
\bjournal{Bioinformatics}
\bvolume{21}
\bpages{1764--1775}.
\bid{doi={10.1093/bioinformatics/bti254}, issn={1367-4803}, pii={bti254},
pmid={15673564}}
\end{barticle}
%
\endbibitem

%b26 ###
\bibitem[\protect\citeauthoryear{Morris et~al.}{2006}]{MorrBrowetal2006}
%
\begin{bincollection}[auto:STB|2011-03-03|12:04:44]
\bauthor{\bsnm{Morris},~\bfnm{J.~S.}\binits{J.~S.}},
\bauthor{\bsnm{Brown},~\bfnm{P.~J.}\binits{P.~J.}},
\bauthor{\bsnm{Baggerly},~\bfnm{K.~A.}\binits{K.~A.}} \AND
\bauthor{\bsnm{Coombes},~\bfnm{K.~R.}\binits{K.~R.}}
(\byear{2006}).
\btitle{Analysis of mass spectrometry data using Bayesian wavelet-based
functional mixed models}.
In \bbooktitle{Bayesian Inference for Gene Expression and Proteomics}
(\beditor{\bfnm{Kim-Ahn}\binits{K.-A.}~\bsnm{Do}},
\beditor{\bfnm{Peter}\binits{P.}~\bsnm{M{\"u}ller}} \AND
\beditor{\bfnm{Marina}\binits{M.}~\bsnm{Vannucci}}, eds.)
\bpages{Chapter~14, 269--292}.
\bpublisher{Cambridge Univ. Press}, \baddress{Cambridge}.
\end{bincollection}
%
\endbibitem

%b27 ###
\bibitem[\protect\citeauthoryear{Morris et~al.}{2008}]{MorrBrowHerretal2008}
%
\begin{barticle}[mr]
\bauthor{\bsnm{Morris},~\bfnm{Jeffrey~S.}\binits{J.~S.}},
\bauthor{\bsnm{Brown},~\bfnm{Philip~J.}\binits{P.~J.}},
\bauthor{\bsnm{Herrick},~\bfnm{Richard~C.}\binits{R.~C.}},
\bauthor{\bsnm{Baggerly},~\bfnm{Keith~A.}\binits{K.~A.}} \AND
\bauthor{\bsnm{Coombes},~\bfnm{Kevin~R.}\binits{K.~R.}}
(\byear{2008}).
\btitle{Bayesian analysis of mass spectrometry proteomic data using
wavelet-based functional mixed models}.
\bjournal{Biometrics}
\bvolume{64}
\bpages{479--489, 667}.
\bid{doi={10.1111/j.1541-0420.2007.00895.x}, issn={0006-341X}, mr={2432418}}
\end{barticle}
%
\endbibitem

%b28 ###
\bibitem[\protect\citeauthoryear{M{\"u}ller et~al.}{2010}]{MullBaggetal2010}
%
\begin{bincollection}[auto:STB|2011-03-03|12:04:44]
\bauthor{\bsnm{M{\"u}ller},~\bfnm{P.}\binits{P.}},
\bauthor{\bsnm{Baggerly},~\bfnm{K.~A.}\binits{K.~A.}},
\bauthor{\bsnm{Do},~\bfnm{K.-A.}\binits{K.-A.}} \AND
\bauthor{\bsnm{Bandyopadhyay},~\bfnm{R.}\binits{R.}}
(\byear{2010}).
\btitle{A Bayesian mixture model for protein biomarker discovery}.
In \bbooktitle{Bayesian Modeling in Bioinformatics}
(\beditor{D.~K. Dey},
\beditor{S.~Ghosh} \AND
\beditor{B.~K. Mallick}, eds.).
\bpublisher{Chapman \& Hall/CRC Press}, \baddress{Boca Raton, FL}.
\end{bincollection}
%
\endbibitem

%b29 ###
\bibitem[\protect\citeauthoryear{Nguyen et~al.}{2010}]{Nguyetal2010}
%
\begin{barticle}[pbm]
\bauthor{\bsnm{Nguyen},~\bfnm{Nha}\binits{N.}},
\bauthor{\bsnm{Huang},~\bfnm{Heng}\binits{H.}},
\bauthor{\bsnm{Oraintara},~\bfnm{Soontorn}\binits{S.}} \AND
\bauthor{\bsnm{Vo},~\bfnm{An}\binits{A.}}
(\byear{2010}).
\btitle{Mass spectrometry data processing using zero-crossing lines in
multi-scale of Gaussian derivative wavelet}.
\bjournal{Bioinformatics}
\bvolume{26}
\bpages{i659--i665}.
\bid{doi={10.1093/bioinformatics/btq397}, issn={1367-4811}, pii={btq397},
pmcid={2935426}, pmid={20823336}}
\end{barticle}
%
\endbibitem

%b30 ###
\bibitem[\protect\citeauthoryear{R Development Core Team}{2010}]{R2010}
%
\begin{bmisc}[auto:STB|2011-03-03|12:04:44]
\borganization{R Development Core Team}
(\byear{2010}).
\bhowpublished{\texttt{R}: \textit{A Language and Environment for Statistical
Computing}. R Foundation for Statistical Computing, Vienna}.
\end{bmisc}
%
\endbibitem

%b31 ###
\bibitem[\protect\citeauthoryear{Siuzdak}{2003}]{Siuz2003}
%
\begin{bbook}[auto:STB|2011-03-03|12:04:44]
\bauthor{\bsnm{Siuzdak},~\bfnm{G.}\binits{G.}}
(\byear{2003}).
\btitle{The Expanding Role of Mass Spectrometry in Biotechnology.}
\bpublisher{MCC Press}, \baddress{San Diego, CA}.
\end{bbook}
%
\endbibitem

%b32 ###
\bibitem[\protect\citeauthoryear{Tibshirani et~al.}{2004}]{Tibsetal2004}
%
\begin{barticle}[pbm]
\bauthor{\bsnm{Tibshirani},~\bfnm{Robert}\binits{R.}},
\bauthor{\bsnm{Hastie},~\bfnm{Trevor}\binits{T.}},
\bauthor{\bsnm{Narasimhan},~\bfnm{Balasubramanian}\binits{B.}},
\bauthor{\bsnm{Soltys},~\bfnm{Scott}\binits{S.}},
\bauthor{\bsnm{Shi},~\bfnm{Gongyi}\binits{G.}},
\bauthor{\bsnm{Koong},~\bfnm{Albert}\binits{A.}} \AND
\mbox{\bauthor{\bsnm{Le}, \bfnm{Quynh-Thu}\binits{Q.-T.}}}
(\byear{2004}).
\btitle{Sample classification from protein mass spectrometry, by 'peak
probability contrasts'}.
\bjournal{Bioinformatics}
\bvolume{20}
\bpages{3034--3044}.
\bid{doi={10.1093/bioinformatics/bth357}, issn={1367-4803}, pii={bth357},
pmid={15226172}}
\end{barticle}
%
\endbibitem

%b33 ###
\bibitem[\protect\citeauthoryear{Vidakovic}{1999}]{Vida1999}
%
\begin{bbook}[mr]
\bauthor{\bsnm{Vidakovic},~\bfnm{Brani}\binits{B.}}
(\byear{1999}).
\btitle{Statistical Modeling by Wavelets}.
\bpublisher{Wiley}, \baddress{New York}.
\bid{doi={10.1002/9780470317020}, mr={1681904}}
\end{bbook}
%
\endbibitem

%b34 ###
\bibitem[\protect\citeauthoryear{Wand and Jones}{1995}]{WandJone1995}
%
\begin{bbook}[mr]
\bauthor{\bsnm{Wand},~\bfnm{M.~P.}\binits{M.~P.}} \AND
\bauthor{\bsnm{Jones},~\bfnm{M.~C.}\binits{M.~C.}}
(\byear{1995}).
\btitle{Kernel Smoothing}.
\bseries{Monographs on Statistics and Applied Probability}
\bvolume{60}.
\bpublisher{Chapman \& Hall}, \baddress{London}.
\bid{mr={1319818}}
\end{bbook}
%
\endbibitem

%b35 ###
\bibitem[\protect\citeauthoryear{Wang, Ray and
Mallick}{2007}]{WangRayMall2007}
%
\begin{barticle}[mr]
\bauthor{\bsnm{Wang},~\bfnm{Xiaohui}\binits{X.}},
\bauthor{\bsnm{Ray},~\bfnm{Shubhankar}\binits{S.}} \AND
\bauthor{\bsnm{Mallick},~\bfnm{Bani~K.}\binits{B.~K.}}
(\byear{2007}).
\btitle{Bayesian curve classification using wavelets}.
\bjournal{J.~Amer. Statist. Assoc.}
\bvolume{102}
\bpages{962--973}.
\bid{doi={10.1198/016214507000000455}, issn={0162-1459}, mr={2354408}}
\end{barticle}
%
\endbibitem

%b36 ###
\bibitem[\protect\citeauthoryear{Wang et~al.}{2003}]{WangHowaCampEtal2003}
%
\begin{barticle}[auto:STB|2011-03-03|12:04:44]
\bauthor{\bsnm{Wang},~\bfnm{M.~Z.}\binits{M.~Z.}},
\bauthor{\bsnm{Howard},~\bfnm{B.~A.}\binits{B.~A.}},
\bauthor{\bsnm{Campa},~\bfnm{M.~J.}\binits{M.~J.}},
\bauthor{\bsnm{Patz},~\bfnm{E.~F.}\binits{E.~F.} \bsuffix{Jr.}}
\AND
\bauthor{\bsnm{Fitzgerald},~\bfnm{M.~C.}\binits{M.~C.}}
(\byear{2003}).
\btitle{Analysis of human serum proteins by liquid phase iso-electric focusing
and matrix-assisted laser desorption/ionization mass spectrometry}.
\bjournal{Proteomics}
\bvolume{3}
\bpages{1661--1666}.
\end{barticle}
%
\endbibitem

%b37 ###
\bibitem[\protect\citeauthoryear{Wolpert, Clyde and
Tu}{2011}]{WolpClydTu2006}
%
\begin{bmisc}[auto:STB|2011-03-03|12:04:44]
\bauthor{\bsnm{Wolpert},~\bfnm{R.~L.}\binits{R.~L.}},
\bauthor{\bsnm{Clyde},~\bfnm{M.~A.}\binits{M.~A.}} \AND
\bauthor{\bsnm{Tu},~\bfnm{C.}\binits{C.}}
(\byear{2011}).
\bhowpublished{Stochastic expansions using continuous dictionaries: L\'evy
adaptive regression kernels. \textit{Ann. Statist}. To appear.}
\end{bmisc}
%
\endbibitem

%b38 ###
\bibitem[\protect\citeauthoryear{Wolpert and Ickstadt}{2004}]{WolpIcks2004}
%
\begin{barticle}[mr]
\bauthor{\bsnm{Wolpert},~\bfnm{Robert~L.}\binits{R.~L.}} \AND
\bauthor{\bsnm{Ickstadt},~\bfnm{Katja}\binits{K.}}
(\byear{2004}).
\btitle{Reflecting uncertainty in inverse problems: A~{B}ayesian
solution using
{L}\'evy processes}.
\bjournal{Inverse Problems}
\bvolume{20}
\bpages{1759--1771}.
\bid{doi={10.1088/0266-5611/20/6/004}, issn={0266-5611}, mr={2107235}}
\end{barticle}
%
\endbibitem

%b39 ###
\bibitem[\protect\citeauthoryear{Yasui et~al.}{2003}]{YasuMcLeetal2003}
%
\begin{barticle}[pbm]
\bauthor{\bsnm{Yasui},~\bfnm{Yutaka}\binits{Y.}},
\bauthor{\bsnm{McLerran},~\bfnm{Dale}\binits{D.}},
\bauthor{\bsnm{Adam},~\bfnm{Bao-Ling}\binits{B.-L.}},
\bauthor{\bsnm{Winget},~\bfnm{Marcy}\binits{M.}},
\bauthor{\bsnm{Thornquist},~\bfnm{Mark}\binits{M.}} \AND
\bauthor{\bsnm{Feng},~\bfnm{Ziding}\binits{Z.}}
(\byear{2003}).
\btitle{An automated peak identification/calibration procedure for
high-dimensional protein measures from mass spectrometers}.
\bjournal{J. Biomed. Biotechnol.}
\bvolume{2003}
\bpages{242--248}.
\bid{doi={10.1155/S111072430320927X}, issn={1110-7251},
pii={S111072430320927X}, pmcid={514270}, pmid={14615632}}
\end{barticle}
%
\endbibitem

%b40 ###
\bibitem[\protect\citeauthoryear{Zhigilei and Garrison}{1998}]{ZhiGarr1998}
%
\begin{barticle}[auto:STB|2011-03-03|12:04:44]
\bauthor{\bsnm{Zhigilei},~\bfnm{L.~V.}\binits{L.~V.}} \AND
\bauthor{\bsnm{Garrison},~\bfnm{B.~J.}\binits{B.~J.}}
(\byear{1998}).
\btitle{Velocity distributions of analyte molecules in matrix assisted laser
desorption from computer simulations}.
\bjournal{Rapid Communications in Mass Spectrometry}
\bvolume{12}
\bpages{1273--1277}.
\end{barticle}
%
\endbibitem

\end{thebibliography}
\end{document}